\documentclass[12pt,preprint]{aastex}

\newcommand{\teff}{$T_{\rm eff}$}
\newcommand{\tc}{$T_{\mathrm{c}}$}
\newcommand{\swan}{C$_{2}$}
\newcommand{\kep}{{\it Kepler}}

\begin{document}

\title{DETAILED ABUNDANCES OF STARS WITH SMALL PLANETS DISCOVERED BY {\it KEPLER} I: The First Sample\altaffilmark{*}}

\altaffiltext{*}{Some of the data presented herein were obtained at the W.M. Keck Observatory, which is operated as a scientific partnership among the California Institute of Technology, the University of California and the National Aeronautics and Space Administration. The Observatory was made possible by the generous financial support of the W.M. Keck Foundation.}

\author{Simon C. Schuler\altaffilmark{1,11}, Zachary A. Vaz\altaffilmark{1}, Orlando J. Katime Santrich\altaffilmark{2}, Katia Cunha\altaffilmark{2,3}, Verne V. Smith\altaffilmark{2,4}, Jeremy R. King\altaffilmark{5}, Johanna K. Teske\altaffilmark{6,7}, Luan Ghezzi\altaffilmark{8}, Steve B. Howell\altaffilmark{9,11}, AND Howard Isaacson\altaffilmark{10}}

\affil{
	\altaffiltext{1}{University of Tampa, Tampa, FL, 33606  USA; sschuler@ut.edu, zachary.vaz@spartans.ut.edu}
	\altaffiltext{2}{Observat{\'o}rio Nacional, Rio de Janeiro, RJ, Brazil; osantrich@on.br, kcunha@noao.edu, vsmith@noao.edu}
	\altaffiltext{3}{Steward Observatory, University of Arizona, Tucson, AZ  85721  USA}
	\altaffiltext{4}{National Optical Astronomy Observatory, Tucson, AZ, 85719  USA}
	\altaffiltext{5}{Department of Physics and Astronomy, Clemson University, Clemson, SC 29634  USA; jking2@clemson.edu}
	\altaffiltext{6}{Department of Terrestrial Magnetism, Carnegie Institution of Washington, Washington, DC 20015  USA; jteske@carnegiescience.edu}
	\altaffiltext{7}{Carnegie Origins Fellow, joint appointment between Carnegie DTM/OCIW}
	\altaffiltext{8}{Harvard-Smithsonian Center for Astrophysics, Cambridge, MA 02138  USA; lghezzi@cfa.harvard.edu}
	\altaffiltext{9}{NASA Ames Research Center, Moffett Field, CA  94035  USA; steve.b.howell@nasa.gov}
	\altaffiltext{10}{Department of Astronomy, University of California, Berkeley, CA 94720  USA; hisaacson@berkeley.edu} 
	\altaffiltext{11}{Visiting Astronomer, Kitt Peak National Observatory, National Optical Astronomy Observatory, which is operated by the Association of Universities for Research in Astronomy (AURA) under cooperative agreement with the National Science Foundation.}
	}

\begin{abstract}
We present newly derived stellar parameters and the detailed abundances of 19 elements of seven stars with small planets discovered by NASA's \kep\ Mission. Each star save one has at least one planet with a radius $\leq 1.6 \; R_{\earth}$, suggesting a primarily rocky composition. The stellar parameters and abundances are derived from high signal-to-noise ratio, high-resolution echelle spectroscopy obtained with the 10-m Keck I telescope and HIRES spectrometer using standard spectroscopic techniques. The metallicities of the seven stars range from -0.32 dex to +0.13 dex, with an average metallicity that is subsolar, supporting previous suggestions that, unlike Jupiter-type giant planets, small planets do not form preferentially around metal-rich stars. The abundances of elements other than iron are in line with a population of Galactic disk stars, and despite our modest sample size, we find hints that the compositions of stars with small planets are similar to stars without known planets and with Neptune-size planets, but not to those of stars with giant planets. This suggests that the formation of small planets does not require exceptional host-star compositions and that small planets may be ubiquitous in the Galaxy.  We compare our derived abundances (which have typical uncertainties of $\lesssim 0.04$ dex) to the condensation temperature of the elements; a correlation between the two has been suggested as a possible signature of rocky planet formation. None of the stars demonstrate the putative rocky planet signature, despite at least three of the stars having rocky planets estimated to contain enough refractory material to produce the signature, if real. More detailed abundance analyses of stars known to host small planets are needed to verify our results and place ever more stringent constraints on planet formation models.
\end{abstract}


\section{INTRODUCTION}
\label{s:intro}
One of the greatest surprises to emerge from the growing list of known exoplanets is the incredible diversity in both the planets themselves and in the architectures of the planetary systems. Discovered planets range in size from the sub-mercury Kepler-37b \citep{2013Natur.494..452B} to objects that straddle the boundary between planets and brown dwarfs, e.g. XO-3b \citep{2008ApJ...677..657J}, with a radius nearly twice that of Jupiter and a mass approaching the deuterium-burning limit of 13 $M_{\rm J}$. Hundreds of stars with single planets having semi-major axes ranging from less than one hundredth of an AU \citep[Kepler-78b,][]{2013ApJ...774...54S} to more than 100 AU \citep[Fomalhaut b,][]{2008Sci...322.1345K} are known, but most systems have two or more planets. While none of the systems discovered to date can truly be called Solar-System analogs, the seven-planet system Kepler-90 \citep{2014ApJ...781...18C,2014ApJ...784...44L} comes closest, with its small planets on orbits interior to its larger, Jupiter-type giant planets. The most incredible systems of all may be those with circumbinary planets, in which the planets orbit both stars of a binary system \citep[e.g., Kepler-16,][]{2011Sci...333.1602D}.

Significant research efforts have been focused on characterizing the discovered planetary systems with the goal of determining planet properties, constraining planet formation models, and providing input for asteroseismic analyses, among others. A critical component of these efforts has been the derivation of the stellar parameters, metallicities, and abundances of numerous elements of the host stars. It is now well established that Jupiter-type giant planets form more readily in metal-rich environments, the so-called \textit{planet-metallicity} correlation \citep[e.g.,][]{2001A&A...373.1019S,2005ApJ...622.1102F,2010ApJ...720.1290G}, with a typical offset of $\sim0.15$ dex in the metallicities found for stars with giant planets and stars without known giant planets. It is now also becoming more well established that stars with small planets do not follow this planet-metallicity correlation and that indeed Neptune-size and smaller planets form at a range of metallicities \citep{2006A&A...447..361U,2008A&A...487..373S,2010ApJ...720.1290G,2012Natur.486..375B,2013ApJ...771..107E}.  The penultimate of these works, \citet{2012Natur.486..375B}, uses the automated spectral fitting routine Stellar Parameter Classification (SPC) to determine the stellar parameters (e.g., \teff, $\log g$, [m/H]\footnotemark[12]) for 152 stars harboring 226 planet candidates (175 of which have radii less than that of Neptune) discovered by NASA's \kep\ Mission. Two main conclusions are reached by the authors.  One, planets with radii less than four Earth radii form around stars with a wide range of metallicities, ranging from -0.6 dex to +0.5 dex.  Two, the average metallicity of stars having planets with radii less than four Earth radii is near solar ($-0.01 \pm 0.02$ dex), which is significantly lower than that of stars with giant planets. Moreover, \citet{2015ApJ...808..187B} recently find that stars that host small planets ($R < 1.7 R_{\earth}$) have the same average metallicity ($-0.02 \pm 0.02$ dex) as stars without detected transiting planets. \citet{2015AJ....149...14W}, though, have suggested the planet-metallicity correlation is universal and does apply to small planets, although at a lower significance than larger Jupiter-type giant planets.

\footnotetext[12]{SPC gives the metallicity ([m/H]) in terms of the average abundance of all elements other than H and He, as opposed to the metallicity ([Fe/H]) based on the Fe content of the star, as is more common. Following \citet{2012ApJ...757..161T}, we assume the two are equivalent throughout the paper.}

\citet{2014Natur.509..593B} find that, based on a homogeneous analysis of more than 400 stars hosting 600 planets discovered by \kep, planets may be categorized into three distinct populations defined by the metallicities of the host stars. The authors suggest that small, rocky planets ($R < 1.7 R_{\earth}$) are found around stars with an average metallicity of [m/H] $= -0.02 \pm 0.02$, the slightly larger gas dwarf planets ($1.7 R_{\earth} < R < 3.9 R_{\earth}$) around stars with an average metallicity of [m/H] $= +0.05 \pm 0.01$, and the gas giant planets ($R > 3.9 R_{\earth}$) around stars with an average metallicity of [m/H] $= +0.18 \pm 0.02$. Their main conclusion is that host star metallicity, and thus by proxy the protoplanetary disk metallicity, is a primary determinant of the final masses and compositions of small planets. In contrast, \citet{2015ApJ...799L..26S} argues that there is no evidence for such discrete planet formation regimes, citing possible shortcomings in the \citet{2014Natur.509..593B} analysis, and that it is more likely that a continuum of planet sizes from 1 $R_{\earth}$ to 4 $R_{\earth}$ form independent of host star metallicity. Therefore, Schlaufman's main conclusion is that host star and protoplanetary disk metallicity is unlikely to have a primary influence on the final masses and compositions of small planets.

Potentially more interesting is what the abundances of individual elements can inform us about the formation of planets, as recently demonstrated by \citet{2015A&A...580L..13S} who has shown that host star abundances can be used to constrain the compositions of orbiting rocky planets. For instance, the abundance of Li has been found to be depleted in stars hosting giant planets \citep[e.g.][]{2009Natur.462..189I,2014A&A...562A..92D}, although there is some disagreement on whether the depleted abundances are due to planet formation or not \citep[e.g.][]{2010ApJ...724..154G,2012ApJ...756...46R}. \citet{2012A&A...547A..36A} find that the abundances of $\alpha$-elements are enhanced in metal-poor stars with rocky planets, suggesting that small planet formation is more efficient in environments enhanced with metals other than Fe. It also has been suggested that the formation of small, rocky planets may leave a chemical imprint in their host stars. \citet{2009ApJ...704L..66M} reported that the refractory elements, those elements with condensation temperatures (\tc\ ) greater than 900 K, are depleted in the solar photosphere compared to nearby solar twins and analogs, and that the depletions are a function of \tc, i.e., elements with higher \tc\ are increasingly more depleted. The authors suggest that the ``missing'' refractory elements are contained in the four rocky planets of our Solar System. \citet{2009A&A...508L..17R} were the first to look for this rocky planet signature in stars other than the Sun by deriving detailed abundances of 64 solar twins and analogs, and measuring the slope of a linear least-square fit to the refractory abundances as a function of \tc. They found that $\simeq 15\%$ of the stars with [Fe/H] $\leq 0.1$ have slopes that are consistent with depleted refractory elements.

Numerous studies have since investigated the putative rocky planet signature by analyzing additional solar twins and analogs with and without known planets, stars with known Jupiter-type giant, Neptune-type, and super-Earth planets, and late F-type dwarf stars. In general, most studies find stars with slopes indicative of depleted refractories and stars with slopes that are not, but there are substantive differences in the interpretations of the observed slopes.  \citet{2010A&A...521A..33R,2012A&A...543A..29M}, and \citet{2014A&A...561A...7R,2015ApJ...808...13R} argue that refractory depletions that they observe are in fact due to rocky planet formation. \citet{2010ApJ...720.1592G,2013A&A...552A...6G} find no evidence that [x/Fe]-\tc\ trends are related to the presence of rocky planets. \citet{2011A&A...528A..85O} find an abundance pattern similar to that of the Sun in a solar twin in the M67 open cluster and suggest the depleted refractories in both stars may be due to radiative dust cleansing by nearby luminous stars during their formation; a follow-up study of additional M67 stars reaches the same conclusion \citep{2014A&A...562A.102O}. \citet{2011ApJ...732...55S} show that Galactic chemical evolution may be responsible for the depleted refractories they find in a sample of stars with Jupiter-type giant planets. \citet{2014A&A...564L..15A} conclude that the age and Galactic birthplace of a star determine its chemical composition and suggest that planet formation does not deplete refractory elements from host stars. \citet{2015A&A...579A..20M} studied the abundances of stars with and without circumstellar debris discs and found no differences in the compositions of the two; furthermore, they find that stars with debris discs and low-mass planets have compositions similar to stars with no detected planets. Finally, \citet{2015A&A...579A..52N} derived detailed abundances for a sample of solar twins and concludes that the interpretation of [x/Fe]-\tc\ trends is complicated because of the convoluted effects of planet formation or dust-gas segregation, stellar age, and Galactic chemical evolution. \citet{2015ApJ...804...40G} has also recently argued that dust-gas segregation is a plausible explanation for the [x/Fe]-\tc\ trends.

As the discord among these various studies demonstrates, it is yet unclear if [x/Fe]-\tc\ trends are indicative of rocky planet formation. However, it is important to emphasize that aside from the fine analysis of the Sun by \citet{2009ApJ...704L..66M}, none of the studies discussed above have included stars with known Earth-like rocky planets. We have undertaken a concerted effort to address this issue by targeting stars with known rocky planets discovered by \kep. Here, as a first step, we present the results of a detailed abundance analysis of seven \kep\ stars, six of which have at least one planet with a radius less than or approximately equal to $1.6 \; R_{\earth}$, and thus their detailed abundances presented here are important contributions to the ongoing efforts to characterize small planet host stars. We compare the abundances of each star to a general population of Galactic disk stars in order to place them into a larger context. We also analyze the refractory element abundances as a function of \tc\ to test the hypothesis that rocky planet formation leaves an observable signature in the composition of the host stars.

\section{OBSERVATIONS AND DATA REDUCTIONS}
\label{s:obs}
We have analyzed high-resolution, high-signal-to-noise (S/N) ratio spectra of seven stars, each of which has at least one confirmed small planet discovered by \kep. The stars are:

\noindent{\bf Kepler-20}: a G8 V\footnotemark[13] dwarf with an age of $8.8^{+4.7}_{-2.7}$ Gyr and host of five known planets \citep{2012ApJ...749...15G,2012Natur.482..195F}.  Kepler-20e and Kepler-20f have radii of $0.868^{+0.074}_{-0.096} \; R_{\earth}$ and $1.03^{+0.10}_{-0.13} \; R_{\earth}$, respectively, and thus are considered Earth-size planets.  Kepler-20b has a radius of $1.91^{+0.12}_{-0.21} \; R_{\earth}$ and is classified as a super-Earth.  The remaining two planets, Kepler-20c and Kepler-20d, have been labeled as sub-Neptunes ($R_{\rm{Neptune}} = 3.88 \; R_{\earth}$) with radii of $3.07^{+0.20}_{-0.31} \; R_{\earth}$ and $2.75^{+0.17}_{-0.30} \; R_{\earth}$, respectively.  Only the masses of Kepler-20b ($M_{\mathrm{b}}=8.7^{+2.1}_{-2.2} \; M_{\earth}$) and Kepler-20c ($M_{\mathrm{c}}=16.1^{+3.3}_{-3.7} \; M_{\earth}$) have been determined definitively \citep{2012ApJ...749...15G}.

\noindent{\bf Kepler-21}: a F6 IV\footnotemark[13] subgiant with an age of $2.84 \pm 0.34$ Gyr and host of one known planet \citep{2012ApJ...746..123H}.  Kepler-21b has a radius of $1.64 \pm 0.04 \; R_{\earth}$ and is classified as a super-Earth.  A definitive mass has yet to be determined for this planet.

\footnotetext[13]{http://simbad.u-strasbg.fr/simbad/sim-fid}

\noindent{\bf Kepler-22}: a G5 V\footnotemark[14] dwarf without a well-determined age and host to one known planet \citep{2012ApJ...745..120B}.  Kepler-22b has a radius of $2.38 \pm 0.13 \; R_{\earth}$ and is classified as a sub-Neptune.  A definitive mass has yet to be determined for this planet.

\footnotetext[14]{http://exoplanetarchive.ipac.caltech.edu/}

\noindent{\bf Kepler-37}: a main-sequence (MS) dwarf of undetermined spectral type with an age of 5.66 Gyr \citep[][no uncertainty provided]{2014ApJS..210...20M} and host to three known planets \citep{2013Natur.494..452B}.  Kepler-37b is a sub-Mercury-size planet with a radius of $0.303^{+0.053}_{-0.073}
\; R_{\earth}$.  Kepler-37c is a Earth-size planet with a radius of $0.742^{+0.065}_{-0.083}\; R_{\earth}$, and Kepler-37d is a super-Earth with a radius of $1.99^{+0.11}_{-0.14}\; R_{\earth}$.  Definitive masses have yet to be determined, but \citet{2014ApJS..210...20M} estimate masses of $M_{\mathrm{b}}=2.78 \pm 3.7 \; M_{\earth}$, $M_{\mathrm{c}}=3.35 \pm 4.0 \; M_{\earth}$, and $M_{\mathrm{d}}=1.87 \pm 9.08 \; M_{\earth}$ for the three planets using a posterior distribution based on a Markov-Chain-Monte-Carlo routine.

\noindent{\bf Kepler-68}: a MS dwarf of undetermined spectral type with an age of $6.3 \pm 1.7$ Gyr and host to three known planets \citep{2013ApJ...766...40G}.  The two inner planets, Kepler-68b and Kepler-68c, are a sub-Neptune with a radius of $2.31^{+0.06}_{-0.09}\; R_{\earth}$ and a Earth-size planet with a radius of $0.953^{+0.037}_{-0.042}\; R_{\earth}$, respectively.  Follow-up RV measurements of Kepler-68 revealed a third, Jupiter-type giant planet, Kepler-68d, orbiting beyond the two inner planets.  The initial RV measurements yielded a mass of $M_{\mathrm{b}} = 8.3^{+2.2}_{-2.4} \; M_{\earth}$ for Kepler-68b and a minimum mass of $M\sin i = 0.947 \pm 0.035 \; M_{J}$ for Kepler-68d \citep{2013ApJ...766...40G}. Analysis of additional RV data refined these values to $M_{\mathrm{b}} = 5.97 \pm 1.7 \; M_{\earth}$ and $M\sin i = 0.84 \pm 0.006 \; M_{J}$, respectively \citep{2014ApJS..210...20M}. A definitive mass has yet to be determined for Kepler-68c.

\noindent{\bf Kepler-100}: a MS dwarf of undetermined spectral type with an age of 6.5 Gyr and host to three known planets \citep{2014ApJS..210...20M}.  Kepler-100b and Kepler-100d have radii of $1.32 \pm 0.04 \; R_{\earth}$ and $1.61 \pm 0.05 \; R_{\earth}$, respectively, and thus are classified as super-Earths.  Kepler-100c, a sub-Neptune, has a radius of $2.20 \pm 0.05 \; R_{\earth}$. Definitive masses have yet to be determined, but \citet{2014ApJS..210...20M} estimate masses of $M_{\mathrm{b}} = 7.34 \pm 3.2 \; M_{\earth}$ and $M_{\mathrm{c}} = 0.85 \pm 4.0 \; M_{\earth}$ for Kepler-100b and Kepler-100c, respectively.

\noindent{\bf Kepler-130}: a MS dwarf of undetermined spectral type without a well-determined age and host to three known planets \citep{2014ApJ...784...45R}.  Kepler-130b is an Earth-size planet with a radius of $1.02 \pm 0.04 \; R_{\earth}$, and Kepler-130d is a super-Earth with a radius of $1.64 \pm 0.16 \; R_{\earth}$. Kepler-130c is classified as a sub-Neptune with a radius of $2.81 \pm 0.09 \; R_{\earth}$. Definitive masses have yet to be determined for these planets.

The spectra of these stars were obtained as part of the \kep\ Follow-up Observing Program (KFOP), an effort orchestrated by \kep\ science team members to confirm planetary candidates discovered by \kep\ and to characterize the planetary systems and their host stars.  The 10-m Keck I telescope and High Resolution Echelle Spectrometer \citep[HIRES;][]{1994SPIE.2198..362V} are being utilized for precise RV measurements of high-priority KOIs in order to confirm bona fide planets and to place constraints on their masses.  Precise RVs of the host stars are made possible by the insertion of an iodine cell into the spectrometer's light path, superimposing the iodine spectrum on the stellar spectrum and providing a stable spectral reference against which subtle shifts in the stellar spectrum can be measured.  The iodine reference spectrum is dense with absorption lines from approximately 5000 -- 6300 {\AA} \citep{1992PASP..104..270M} and makes the stellar spectrum unusable in this region for abundance analysis.  Instead, we use the template spectra-- spectra of the target stars taken with the same instrumental setup but without the iodine cell-- for our abundance analysis.  Moreover, only stars with spectra having S/N ratios per pixel greater than 150 were considered for study.  The instrumental setup, observational procedure, and data reduction used for the KFOP targets are the same as those used by the California Planet Search survey \citep{2008PhST..130a4001M}.  The KFOP spectra are characterized by a spectral resolution of $R = \lambda / \Delta \lambda = 50,000$ and span 3650 -- 7950 {\AA} with incomplete coverage in the reddest orders.

Additional observations of Kepler-21 were made independently with Keck/HIRES and the 4-m Mayall telescope and echelle spectrograph at Kitt Peak National Observatory (KPNO) in UT 2011.  The Keck/HIRES observations were made with a different instrumental setup than those of the KFOP.  The setup included a cross-disperser angle of 0.102, an echelle angle of 0.000, the C1 decker, and a projected slit width of 0.86''.  This setup provided a nominal resolution of $R = 50,000$ and a wavelength coverage from 3750 to 8170 {\AA} that is incomplete in the reddest orders.  These Keck/HIRES spectra were reduced with the MAKEE\footnotemark[15] data reduction package, which was developed specifically for HIRES data.  MAKEE is an automated pipeline that carries out the typical reduction procedures, including bias correction, flat fielding, order extraction, and wavelength calibration.

\footnotetext[15]{http://spider.ipac.caltech.edu/staff/tab/makee/index.html}

The KPNO spectrum was obtained during engineering time with the 4-m Mayall telescope.  The echelle spectrograph was setup with the $58.5\, \mathrm{g\, mm}^{-1}$ echelle grating and $226-1\, \mathrm{g\, mm}^{-1}$ cross disperser, resulting in a wavelength coverage of 5270 -- 8060 {\AA} with incomplete spectral coverage in the reddest orders. A projected slit width of 1'' produced a spectral resolution of $R \sim 42,000$.  The T2KA ccd detector with 2048 $\times$ 2048 pixels was used.  Standard routines within the IRAF\footnotemark[16] image processing software were used for the data reductions, which included bias removal, scattered light subtraction, flat fielding, order extraction, and wavelength calibration.  Details of all observations, including the S/N ratios of the spectra, are given in the Observing Log (Table \ref{tab:obs}).  Sample spectra are shown in Figure \ref{fig:spec}.

\footnotetext[16]{IRAF is distributed by the National Optical Astronomy Observatory, which is operated by the Association of Universities for Research in Astronomy, Inc., under cooperative agreement with the National Science Foundation.}

\section{ABUNDANCE ANALYSIS \& RESULTS}
\label{s:abunds}
Our abundance analysis follows that described in \citet{2011ApJ...732...55S} and includes the derivation of stellar parameters (\teff, $\log g$, and microturbulent velocity [$\xi$]) for each star in addition to the abundances of up to 19 elements via an LTE, curve-of-growth analysis.  The abundances have been derived by means of equivalent width (EW) measurements of spectral absorption lines and by the spectral synthesis fitting technique, depending on the line considered.  The lines analyzed are drawn from the line list described in \citet{2011ApJ...732...55S}.  The one-dimensional spectrum analysis package SPECTRE \citep{1987BAAS...19.1129F} was used to measure EWs, primarily utilizing Gaussian profiles; for strong lines (EW $\gtrsim 90$ m{\AA}) with broad wings near the continuum, Voigt profiles were used for the fits.  The 2014 version of the LTE spectral analysis software package MOOG \citep{1973ApJ...184..839S} was used to derive the abundances from the measured EWs and the synthetic fits to the data using model atmospheres interpolated from the extensive grids of Kurucz ATLAS9\footnotemark[17] models with convective overshoot.  Final abundances are reported relative to solar abundances that have been derived in the exact same manner as those for the \kep\ stars using solar spectra obtained with the same instrumental configurations on the Keck and KPNO 4-m telescopes, and adopting the solar parameters \teff$= 5777$ K, $\log g = 4.44$, and $\xi = 1.38 \; \mathrm{km} \; \mathrm{s}^{-1}$.  The one exception is the relative abundances for Kepler-21 derived from the Keck (non-KFOP) spectrum.  No solar spectrum was obtained during the observing run, so abundances derived from the KFOP solar spectrum have been used instead. 

\footnotetext[17]{See http://kurucz/harvard.edu/grids.html}

\subsection{Stellar Parameters}
\label{ss:params}
Stellar parameters for each star have been derived directly from the spectra using excitation and ionization balance of \ion{Fe}{1} and \ion{Fe}{2}.  Initial parameters for each star were taken from the planet validation papers as listed in the Introduction. The parameters were then altered until no correlation existed between the derived [\ion{Fe}{1}/H] and lower excitation potential ($\chi$) and [\ion{Fe}{1}/H] and reduced EW [$\log (\mathrm{EW}/\lambda)$] for each \ion{Fe}{1} line measured, and until the [Fe/H] abundances derived from \ion{Fe}{1} and \ion{Fe}{2} lines were equal to within two significant digits.  Deriving stellar parameters in this way results in unique solutions only if there is no ab initio correlation between $\chi$ and EW of the measured \ion{Fe}{1} lines.  Such correlations did exist for the initial \ion{Fe}{1} line lists for Kepler-20, Kepler-22, and Kepler-37, but the correlations were eliminated by removing two, four, and seven \ion{Fe}{1} lines from their respective line lists.  Uncertainties in \teff\ and $\xi$ are the differences in the adopted parameters and those that result in $1 \sigma$ correlations in the [Fe/H] versus $\chi$ and reduced EW relations, respectively.  The derived surface gravities are sensitive to the Fe abundance derived from both \ion{Fe}{1} and \ion{Fe}{2} lines, with a greater sensitivity to the latter, and thus the uncertainty in $\log g$ is dependent on the uncertainty in the Fe abundances, as more fully described in \citet{2010AJ....140..293B}.

The derived stellar parameters, Fe abundances, and their uncertainties are given in Table \ref{tab:params}. Parameters and Fe abundances have been derived for Kepler-21 from each of the three spectra-- KFOP, Keck (non-KFOP), and KPNO-- obtained for this star.  The results from the three different spectra are indistinguishable within the combined uncertainties, with those from the KFOP and Keck observations in particularly good agreement: $\Delta T_{\mathrm{eff}} = -52$ K, $\Delta \log g = -0.09$, and $\Delta \mathrm{[Fe/H]} = -0.02$.  This is reassuring given the spectra were taken with the same telescope and spectrograph.  The $\log g$ derived from the KPNO spectroscopy deviates from the KFOP and Keck values by 0.28 and 0.19 dex, respectively.  While the differences are approaching statistical significance, they are readily understood given the significantly lower S/N of the KPNO spectrum compared to the Keck and KFOP spectra, as well as the small number of \ion{Fe}{2} lines, and thus higher uncertainty in the mean [\ion{Fe}{2}/H] abundance, measurable in the KPNO spectrum.  The adopted stellar parameters and Fe abundances for Kepler-21 are those derived from the Keck (non-KFOP) spectrum and have been used in the derivation of the other elements. All additional references to the parameters of Kepler-21 will refer to these adopted values.

\subsection{Abundances}
\label{ss:abunds}
The abundances of individual elements that require special attention are described in the subsections that follow.  The stellar and solar EW measurements and the respective absolute abundance for each line of all elements analyzed are given in the Appendix. The final adopted abundances, given relative to the solar, and their uncertainties for each star are provided in Table \ref{tab:abs}.

\subsubsection{Lithium}
Li abundances have been derived from the $\lambda 6707$ Li resonance feature using spectral synthesis. The adopted line list is that from \citet{1997AJ....113.1871K} updated as described in \citet{2012PASP..124..164S}. The synthetic spectra were smoothed using Gaussian profiles with full widths at half maximum (FWHM) determined from weak, unblended absorption lines in the same spectral order as the Li feature. Sample spectra and fits of the $\lambda 6707$ Li region are given in Figure \ref{fig:li}.

\subsubsection{Carbon and Oxygen}
Abundances of the light elements C and O have been derived from a combination of atomic and molecular lines.  For C, we make use of five high-excitation \ion{C}{1} lines and two \swan\ features.  Formation of high-excitation \ion{C}{1} lines is susceptible to non-LTE effects, but abundances derived from the lines measured here have been shown to be largely free of deviations from the LTE approximation for solar-type stars \citep{2005A&A...431..693A,2005PASJ...57...65T}.  The two \swan\ features ($\lambda 5086$ and $\lambda5136$) are blends of multiple components of the \swan\ Swan system; spectral synthesis was used to derive C abundances from these features.  The final adopted C abundance for each star is the mean of the relative abundances derived from the \ion{C}{1} and \swan\ lines, except for Kepler-21, for which only \ion{C}{1} lines were measurable, and for Kepler-37, for which the mean abundance derived from the \ion{C}{1} lines is 0.16 dex higher than that derived from the \swan\ lines. Abundances derived from high-excitation lines, in particular the $\lambda 7775$ high-excitation \ion{O}{1} triplet ($\chi = 9.15$ eV) and $\lambda 6053$ \ion{S}{1} line ($\chi = 7.87$ eV), using one-dimensional LTE analyses have been shown to increase dramatically with decreasing \teff\ in stars with \teff$\lesssim 5400$ K, contrary to canonical NLTE predictions \citep{2004ApJ...602L.117S,2006ApJ...636..432S,2013ApJ...764...78R,2013ApJ...778..132T}. Similar increases in Fe abundances derived from \ion{Fe}{2} lines relative to \ion{Fe}{1} lines are seen in the same temperature regime \citep{2003AJ....125.2085S,2004ApJ...603..697Y,2010PASP..122..766S}. Taken together, the overabundances derived from high-excitation and singly ionized lines in the spectra of cool stars suggest that the electron populations of these energy states are not accurately modeled by our standard LTE analysis. We suspect that the difference in \ion{C}{1} ($\chi = 7.68, 8.65$ eV) and \swan\ abundances of Kepler-37, with \teff $= 5406$ K, is due to the overpopulation of the high-excitation \ion{C}{1} states and thus adopt the \swan\ based abundances for this star. The C abundances of all the stars are given in Table \ref{tab:carb}.

The abundances of O have been derived from the $\lambda 6300$ forbidden [\ion{O}{1}] line and the $\lambda 7775$ high-excitation \ion{O}{1} triplet.  The $\lambda 6300$ line is blended with two isotopic components of a \ion{Ni}{1} transition at 6300.34 {\AA} for which account must be taken when deriving O abundances from this feature.  We use the {\sf blends} driver within the MOOG software package for this purpose.  The necessary input for the abundance derivation includes the EW of the $\lambda 6300$ feature, atomic parameters ($\chi$ and $\log gf$) for both the O and Ni transitions, and a Ni abundance.  The atomic parameters for the [\ion{O}{1}] line are taken from the fine analysis of \citet{2001ApJ...556L..63A}; those for the Ni blend are from the laboratory analysis of \citet{2003ApJ...584L.107J}.  The adopted Ni abundances are those derived here and given below.  Unlike the $\lambda 6300$ forbidden [\ion{O}{1}] line which has been shown to be well described by LTE \citep[e.g.,][]{2003A&A...402..343T}, formation of the high-excitation \ion{O}{1} triplet is sensitive to NLTE effects \citep[e.g.,][]{1991A&A...245L...9K}.  NLTE corrections for the \ion{O}{1} triplet based abundances have been estimated using the analytical formula provided by \citep{2003A&A...402..343T}. We also tested the NLTE corrections of \citet{2009A&A...500.1221F}, which cover a smaller parameter space compared to Takeda et al. The corrections are found to be similar for stars falling within the parameter space of both studies. 

The derived O abundances are provided in Table \ref{tab:oxy}, where it can be seen that the [\ion{O}{1}] and the NLTE-corrected \ion{O}{1} triplet abundances are in good agreement for Kepler-20 and Kepler-21.  Because we consider the [\ion{O}{1}]-based abundances to be more reliable, they are adopted as the final O abundances for these two stars.  The [\ion{O}{1}] line was not measurable in the Kepler-22, Kepler-68, Kepler-100, and Kepler-130 spectra, so the NLTE-corrected triplet-based abundances are adopted for these stars.  We note, however, that the NTLE corrections for Kepler-22 and Kepler-68 are expected to be of the same order as for the Sun \citep{2003A&A...402..343T}, so the LTE and NLTE abundances should not differ significantly.  This is indeed observed, as the LTE and NLTE-based abundances differ by only a few dex in each case (Table \ref{tab:oxy}). Both the [\ion{O}{1}] and \ion{O}{1} triplet features were measurable for Kepler-37, but as described above, the triplet based abundance is likely unreliable due to the overpopulation of high-excitation states in this cool star. We therefore adopt the [\ion{O}{1}] based abundance as the final O abundance for Kepler-37.

\subsubsection{Odd-Z Elements: Sc, V, Mn, and Co}
\label{ss:hyper}
The line profiles of the odd-$Z$ elements Sc, V, Mn, and Co can be affected by hyperfine structure (hfs) in some of the electron transitions of these atoms, resulting in enhanced abundances of these elements if not modeled properly. However, hfs is expected to be significant only in transitions that result in large EWs \citep[$\gtrsim 50$ m{\AA};][]{2000ApJ...537L..57P}. We inspected the EW and relative abundance of each line of these elements for all the stars in our sample, and the abundances of Sc, V, and Co show no indication of being affected by hfs. The Mn abundances of Kepler-20 and Kepler-37, on the other hand, are indicative of hfs effects. For Kepler-20, the EWs of the two measured Mn lines are 55.8 m{\AA} ($\lambda 5400$) and 83.5 m{\AA} ($\lambda 5433$), and the resulting relative abundances are 0.10 dex and 0.29 dex higher, respectively, than the star's mean metallicity. For Kepler-37, the $\lambda 5433$ line, with an EW of 61.6 m{\AA},  shows a slight abundance enhancement of 0.13 dex compared to the $\lambda 5400$ line, which has an EW of only 28.0 m{\AA} and a relative abundance in agreement with the star's mean metallicity.

Corrected Mn abundances of Kepler-20 and Kepler-37 were determined by using spectral synthesis and the hfs components of the $\lambda 5433$ line compiled by \citet{2006ApJ...640..801J} combined with a line list of surrounding features from the Vienna Atomic Line Database \citep[VALD,][]{1999A&AS..138..119K,2000BaltA...9..590K,1997BaltA...6..244R,1995A&AS..112..525P}. To achieve consistent relative abundances, the $\lambda 5433$ line was also synthesized for the Sun. The hfs-corrected abundance for Kepler-20, [Mn/H] $= +0.04$, is 0.31 dex lower than the non-corrected value and 0.12 dex lower than the abundance derived from the $\lambda 5400$ line. The hfs-corrected abundance only is adopted for Kepler-20. For Kepler-37, the hfs-corrected abundance, [Mn/H] $= -0.35$, is only 0.05 dex lower than the non-corrected abundance, signifying that it is only slightly affected by hfs. Nonetheless, we adopt this hfs-corrected value and combine it with that of the $\lambda 5400$ line, which is not expected to be affected by hfs, to obtain the adopted Mn abundance ([Mn/H] $= -0.39$) for Kepler-37.

\subsubsection{Abundance Uncertainties}
Uncertainties in the derived stellar parameters (\teff, $\log g$, and $\xi$) and the uncertainty in the mean abundance ($\sigma_{\mu}$) of a given element contribute to the total internal uncertainty in the derived abundance of that element.  For elements whose abundances are based on two or less lines, the total uncertainty is based on the parameter uncertainties alone.  Abundance sensitivities to the stellar parameters were determined for each star by calculating changes in the final abundances due to parameter changes of $\pm 150$ K in \teff, $\pm 0.25$ dex in $\log g$, and $\pm0.30$ km s$^{-1}$ in $\xi$; representative sensitivities are given in Table \ref{tab:sensi}.  Abundance uncertainties to each parameter are then calculated by scaling the abundance sensitivities by the uncertainty in the respective stellar parameter.  The final total internal uncertainty is the quadratic sum of the individual uncertainties due to the stellar parameters and $\sigma_{\mu}$, for those elements with abundances derived from more than two lines.

\section{DISCUSSION}
\label{s:disc}
\subsection{Comparison to Previous Results}
\label{ss:disc_previous}
Accurately derived stellar parameters are essential for the characterization of exoplanets and as asteroseismic inputs. \kep's incredible success at finding planetary candidates during its primary mission has dramatically increased the number of stars for which effective temperatures, surface gravities, and metallicities are needed in short order, so the size and equilibrium temperature of the planet(s) can be accurately determined. Automated (spectrum fitting) and semi-automated (spectral line synthesis) routines, such as Spectroscopy Made Easy \citep[SME;][]{1996A&AS..118..595V}, SPC \citep{2012Natur.486..375B}, Automatic Routine for line Equivalent widths in stellar Spectra \citep[ARES;][]{2007A&A...469..783S}, and Versatile Wavelength Analysis \citep[VWA;][]{2002A&A...389..345B} to name a few, are generally used to meet this demand. In Table \ref{tab:lit_params} we compare results in the existing literature to ours derived by ``by hand'' to provide a check of those determined using automated or semi-automated routines, and to ensure that the most accurate stellar parameters are being used in planetary characterization and asteroseismic studies. Here we discuss the comparisons for each star.

\noindent{\bf Kepler-20}: Stellar parameters and metallicities have been derived for Kepler-20 by \citet{2012ApJ...749...15G} and \citet{2012Natur.482..195F}, both of which employed SME using the same Keck/HIRES spectrum obtained as part of the KFOP. The two studies find nearly identical parameters and metallicity for this star.  The \teff\ and [Fe/H] of both studies are in agreement with ours within uncertainties, and the surface gravity values are the same at $\log g = 4.44$.

\noindent{\bf Kepler-21}: Numerous spectroscopic and asteroseismic studies have determined the stellar parameters and metallicity of Kepler-21. \citet{2012ApJ...746..123H} carried out an extensive spectroscopic study of the star, analyzing three separate spectra obtained with three different telescopes. Stellar parameters and metallicities were determined from each spectrum using automated spectrum-fitting routines and compared to existing literature values. The final adopted parameters and metallicity were chosen from the accumulated data set. \citet{2013MNRAS.434.1422M} derived two sets of parameters and metallicities using two different routines: the spectrum fitting package ROTFIT {\citep{2003A&A...405..149F} and the semi-automated routine ARES along with MOOG. The two sets of results, with the exception of the metallicities, differ at greater than the $1\sigma$ level, with the ARES/MOOG results being particularly disparate; it is the only analysis suggesting that Kepler-21 has the surface gravity of a MS dwarf rather than a subgiant. \citet{2013ApJ...771..107E} developed a spectrum-fitting routine to analyze low-resolution ($R \sim 3000$) spectra obtained with the KPNO 4-m Mayall telescope and RCSpec long-slit spectrograph. \citet{2012MNRAS.423..122B} used the semi-automated routine VWA to derive \teff, metallicity, and the abundances of 11 other elements, all of which are included in our analysis. The stellar parameters and metallicities of \citet{2012ApJ...746..123H}, \citet{2012MNRAS.423..122B}, and the ROTFIT analysis of \citet{2013MNRAS.434.1422M} are in agreement with ours within uncertainties, while the stellar parameters of \citet{2013ApJ...771..107E}, based on low-resolution spectroscopy, are lower by more than the combined uncertainties. Also, the abundances of the eleven additional elements derived by \citet{2012MNRAS.423..122B} are in good agreement with our results, with all of them agreeing within uncertainties except for the abundance of carbon, our abundance ($-0.02 \pm 0.05$) for which is $+0.18$ dex higher than theirs ($-0.20 \pm -0.07$).

Asteroseismic analyses generally require \teff\ and [Fe/H] as input, and they determine the fundamental stellar properties of mass, radius, and density. With the mass and radius, the surface gravity of a star can be accurately determined \citep[e.g.,][]{2013ARA&A..51..353C}. No less than five different papers report fundamental properties of Kepler-21 based on asteroseismology, and three of the analyses include deriving the input \teff\ and [Fe/H] parameters themselves, either from high-resolution spectroscopy or photometry \citep{2012ApJ...746..123H,2012ApJ...757...99S,2015MNRAS.452.2127S}, one adopts the parameters from the literature \citep{2012ApJ...749..152M}, and one does both \citep{2014ApJS..210....1C}. The derived \teff\ values from these studies range from 5838 K to 6305 K, and the [Fe/H] values from -0.20 to -0.01. Our adopted \teff\ ($6177 \pm 42$ K) and [Fe/H] ($-0.08 \pm 0.07$) fall nicely in the middle of these ranges. More impressive is the tight agreement in the asteroseismically derived surface gravities, which range from 4.0 dex to 4.03 dex, despite the large ranges in \teff\ and [Fe/H]. Our adopted surface gravity, $\log g = 3.99 \pm 0.15$, is in near-perfect agreement with these values.

\noindent{\bf Kepler-22}: Two sets of stellar parameters and metallicities have been derived for Kepler-22 by \citet{2012ApJ...745..120B}. One analysis utilized SME and a Keck/HIRES KFOP spectrum, and the other combined the results of spectrum-fitting analyses of five separate spectra obtained with four different telescopes. \citet{2012ApJ...745..120B} also carried out an asteroseismic analysis, and \citet{2014ApJS..211....2H} calculate a surface gravity based on former's derived mass and radius. Both sets of spectroscopic results are in good agreement with ours, with only \teff\ from the SME analysis deviating from our value by more than the combined uncertainties. Also, our derived surface gravity is 0.13 dex higher, just larger than the combined uncertainty, than the asteroseismically derived value.

\noindent{\bf Kepler-37}: Stellar parameters and metallicity of Kepler-37 are reported in the Kepler-37b discovery paper, \citet{2013Natur.494..452B}. They used SME and SPC to analyze two sets of spectra obtained from two different telescopes as part of the KFOP. After deriving initial parameters and metallicity, asteroseismology was used to determine the star's surface gravity, among other things, and then the surface gravity was held fixed while \teff\ and [Fe/H] were rederived. The final adopted values are the averages of those from the SME and SPC analyses. The \citet{2012Natur.486..375B} values are in excellent agreement with those of Barclay et al., and both sets of results agree with our derived values within the uncertainties. \citet{2013ApJ...771..107E} also analyzed a low-resolution Mayall/RCSpec spectrum of this star. While their surface gravity agrees with ours within the uncertainties, their \teff\ and [Fe/H] are lower than ours by more than the combined uncertainties. Finally, our surface gravity, as well as those of \citet{2013Natur.494..452B} and \citet{2012Natur.486..375B}, is in good agreement with the asteroseismic value of \citet{2015MNRAS.452.2127S}.

\noindent{\bf Kepler-68}: \citet{2013ApJ...766...40G} determined the stellar parameters and metallicity of Kepler-68 using SME and a KFOP spectrum with the surface gravity constrained by an asteroseismic analysis, similar to the analysis of \citet{2013Natur.494..452B}. These results are in good agreement with the \citet{2012Natur.486..375B} values, all of which agree within the uncertainties. The surface gravities are also in agreement with the asteroseismic value determined by \citet{2015MNRAS.452.2127S}. While the results of \citet{2013ApJ...766...40G} agree with ours within the combined uncertainties, the stellar parameters of \citet{2012Natur.486..375B} and the surface gravity of \citet{2015MNRAS.452.2127S} differ from ours by slightly more than the combined uncertainties, with our values larger than those of these two studies. Despite these differences, the metallicities are in very good agreement.

\noindent{\bf Kepler-100}: Initial stellar parameters without uncertainties for Kepler-100 are provided by \citet{2013ApJS..204...24B} based on an SME analysis of a KFOP spectrum. Refined stellar parameters and a metallicity were derived by \citet{2014ApJS..210...20M} using SME and a KFOP spectrum with the surface gravity constrained by an asteroseismic analysis, similar to the analysis of \citet{2013Natur.494..452B}. These updated results are in good agreement with those of \citet{2012Natur.486..375B}. The surface gravities also match well the asteroseismic value of \citet{2015MNRAS.452.2127S}. The results of the latter three studies are in agreement with ours within the combined uncertainties, but those of \citet{2013ApJS..204...24B} differ by more than the uncertainties of our analysis.

\noindent{\bf Kepler-130}: Stellar parameters and metallicity of Kepler-130 have been derived by \citet{2013ApJ...767..127H}  using either SME or SPC and a KFOP spectrum (which routine and the source of the spectrum is not specified for individual stars) with the surface gravity constrained by an asteroseismic analysis, similar to the analysis of \citet{2013Natur.494..452B}. These results are near perfect agreement with those of \citet{2012Natur.486..375B}, and both sets of results agree with ours within the combined uncertainties.

Overall, the statistical agreement between our derived stellar parameters and metallicities of these \kep\ host stars and those derived by numerous other groups using various automated and semi-auotmated routines with {\it high-resolution spectroscopy} is quite good. Only the ARES/MOOG stellar parameters of Kepler-21 \citep{2013MNRAS.434.1422M}, the SME \teff\ of Kepler-22 \citep{2012ApJ...745..120B}, the SPC stellar parameters of Kepler-68 \citep{2012Natur.486..375B}, and the initial SME stellar parameters of Kepler-100 \citep{2013ApJS..204...24B} differ from our values by more than the combined uncertainties. In each case, the metallicity of each analysis agrees with ours. In contrast to the high-resolution spectroscopic analyses, the low-resolution spectroscopic analysis of \citet{2013ApJ...771..107E} produces more disparate results for the \teff\ and $\log g$ of Kepler-21, and the \teff\ and [Fe/H] of Kepler-37, all of which are underestimated at greater than the $2\sigma$ level. Nonetheless, the low-resolution spectroscopic results are considered to be an improvement over the photometrically derived stellar parameters and metallicities given in the \kep\ Input Catalog \citep{2011AJ....142..112B}.

It is also encouraging that our surface gravities statistically agree with the asteroseismic values for four of the six stars in our sample. For the two stars, Kepler-22 and Kepler-68, whose surface gravities are not in agreement, our values are larger at about the $1.7\sigma$ level in each case. It is generally accepted that asteroseismology provides the most accurate and precise surface gravities, because they are largely independent of the input stellar physics \citep{2013ARA&A..51..353C}. It has been shown that, in general, $\log g$ determined spectroscopically by forcing ionization balance are larger than asteroseismically determined values, with the difference increasing with increasing \teff\ \citep[e.g.,][]{2014A&A...572A..95M}. While not all of the stars in our sample follow this general trend, it does agree with what we find for Kepler-22 and Kepler-68. The source of the discrepancy is not yet fully understood and merits further investigation. Nonetheless, our derived \teff\ and [Fe/H] remain robust, because they do not correlate with changes in $\log g$ for standard curve-of-growth analyses \citep{2012ApJ...757..161T,2014A&A...572A..95M}. Furthermore, based on the derived sensitivities (Table \ref{tab:sensi}), the abundances of the other elements should also not be greatly affected by our slightly overestimated $\log g$ for Kepler-22 and Kepler-68.

\subsection{The Abundances of Stars with Small Planets}
\label{ss:disc_abunds}
An immediate conclusion that can be drawn from the abundances given in Table \ref{tab:abs} is that small planets (in this case, $0.30 \; R_{\earth} \lesssim R \lesssim 3.00 \; R_{\earth}$) form in environments with a range of metallicities, from metal-poor ([Fe/H] $\approx -0.30$) to metal-rich ([Fe/H] $\approx +0.15$).  This is in concordance with previous suggestions based on stars with planets classified as sub-Neptunes \citep{2006A&A...447..361U,2008A&A...487..373S,2010ApJ...720.1290G} and, in particular, small planet candidates discovered by \kep\ \citep{2012Natur.486..375B,2013ApJ...771..107E,2015ApJ...808..187B}.  We point out that at the time of this writing that at least 145 of the 226 planet candidates in the \citet{2012Natur.486..375B} sample, including those discussed herein, have since been confirmed as bona fide planets (according to {\it Kepler}'s Table of Confirmed Planets\footnotemark[18]), making more robust their conclusions that small planets form around stars with a wide range of metallicities and that the average metallicity of stars with small planets is approximately solar. Interestingly, the star with the highest metallicity, Kepler-68, is the one star in our sample with a known Jupiter-type giant planet.

\footnotetext[18]{http://kepler.nasa.gov/Mission/discoveries/}

Beyond the bulk metallicity of the host stars, the abundances of 15 of the remaining 18 other elements derived for each star show no deviations from general Galactic abundance distributions. In Figure \ref{fig:gal} we compare the abundances of each element relative to Fe for the stars in our sample to those from studies that have analyzed a large number of stars with and without detected planets in the Galactic disk. Carbon and oxygen abundances are compared to abundances from \citet{2010ApJ...725.2349D}. Additional C abundances are taken from \citet{2006MNRAS.367.1181B} and additonal O abundances from \citet{2014A&A...562A..71B}. The abundances of the remaining elements are compared to those from \citet{2014A&A...562A..71B} and \citet{2012A&A...545A..32A}; the abundances of the latter are for stars with \teff\ between 4900 and 6150 K. As can be seen in the figure, the abundances of each element of the \kep\ planet host stars fall along the Galactic trends, demonstrating that these abundance ratios for this small subset of stars with small planets are typical of the general Galactic disk population. Generally speaking, this implies that the abundances of elements other than Fe scale with Fe in a way that is typical for similar stars in the Galactic disk.

To further demonstrate this, we show in Figures \ref{fig:histo_neptunes} and \ref{fig:histo_jupiters} the distribution of the abundances relative to solar of 13 elements for our sample of Kepler host stars excluding Kepler-68 and the sample of stars with stars with no detected planets, stars with Neptune- and super-Earth-size planets (no Jupiter-type giants), and Jupiter-type giant planets from \citet{2012A&A...545A..32A}. We do not include Kepler-68 in these histograms, because our main interest here is to compare the compositions of stars with small planets to a general stellar population. Kepler-68 is a giant planet (Kepler-68d) host, and given that it has the highest metallicity in our sample, its inclusion would certainly skew the comparisons. Our sample of only six stars prevents us from making firm statistical conclusions based on these comparisons, but they are suggestive of possible correlations. Indeed, the histograms for the Kepler host stars, stars without known planets, and stars with Neptune- and super-Earth-size planets largely overlap, whereas the histograms for the Kepler host stars and stars with Jupiter-type giant planets appear to be offset from each other, with the latter shifted toward higher abundances.

To investigate these comparisons further, two-sample Kolmogorov-Smirnov (KS) tests were carried out for each element in Figures \ref{fig:histo_neptunes} and \ref{fig:histo_jupiters} for the Kepler host star sample and each of the samples of stars without known planets, stars with Neptune- and super-Earth-size planets, and stars with Jupiter-type giant planets.
Again, our small sample size limits the sufficiency of the KS test to provide statistically significant probabilities of the relationships between the different samples of stars, and indeed, none of the tests yielded p-values that rejected the null hypothesis that the samples are drawn from the same parent populations. A much larger sample of detailed abundances of stars with small planets is clearly needed to determine if those abundances are in fact typical of the general Galactic disk population, and lower than those of stars with giant planets, as hinted at in Figures \ref{fig:histo_neptunes} and \ref{fig:histo_jupiters}. If confirmed, this result would complement studies finding that the occurrence rate of small planets around solar-type stars and M dwarfs could be as high as 50\% \citep[e.g.,][]{2012ApJS..201...15H,2013ApJ...766...81F,2013A&A...549A.109B,2014PNAS..11112655M,2015ApJ...807...45D,2015ARA&A..53..409W}, whereas the occurrence rate of Jupiter-type giant planets around these stars is below 10\% \citep[e.g.,][]{2008PASP..120..531C,2015ARA&A..53..409W}, by suggesting that stars do not need enhanced or otherwise distinguishable chemical compositions to form small planets.

We conclude this section with a word on Li. The abundance of Li in stars with planets has been a topic of interest for some time, and there has been a debate on whether planetary hosts exhibit depleted Li abundances relative to stars without known planets for just as long. \citet{2000AJ....119..390G} first suggested that stars with RV-detected planets tend to have Li abundances that are lower than field stars with detected Li based on a sample of seven planet host stars. Soon thereafter, \citet{2000MNRAS.316L..35R} compared the Li abundances of 16 stars with RV-detected planets to a sample of open cluster and field stars with similar ages, evolutionary states, and \teff\ as the planet host stars and found that the Li abundances of the host stars are not different than those of the control sample. Subsequent to these two papers, numerous studies \citep[e.g.,][]{2004A&A...414..601I,2005PASJ...57...45T,2010A&A...512L...5S,2014A&A...570A..21F,2014A&A...562A..92D} argue that stars with RV-detected planets do have Li abundances that are lower on average than stars without detected planets, and numerous studies \citep[e.g.,][]{2006AJ....131.3069L,2010ApJ...724..154G,2010Ap&SS.328..193M,2012ApJ...756...46R} argue that the Li abundances of the two groups are indistinguishable. It is not clear how stars with small rocky (transiting) planets fit into this debate, because the Li abundances of a large enough sample of stars with small planets have yet been determined. Among our sample there are three stars with well-determined Li abundances and four with upper limits (Table \ref{tab:abs}). For the three stars with firm Li detections, their Li abundances are in good agreement with the general Li-\teff\ trends of \citet{2010ApJ...724..154G} and \citet{2015A&A...576A..69D}, and thus the Li abundances of stars with small rocky planets do not appear to be blatantly anomalous. Li abundances of a much larger sample of stars with rocky planets are needed, though, to determine if this is actually the case.

\subsection{Abundances and Condensation Temperatures}
\label{ss:disc_tc}
As mentioned in the Introduction, stellar abundances that correlate with \tc\ may indicate the presence of rocky planets. All of the \kep\ stars in our sample are host to at least one sub-Neptune size or smaller planet, and four have at least one Earth-size planet. RV follow-up observations have, in general, only been able to place upper limits on masses of small planets except in a handful of cases, e.g., Kepler-20b, Kepler-20c, and Kepler-68b, so the densities and refractory element abundances of the majority of small planets have yet to be reliably estimated.  Even with the known densities of Kepler-20b ($\rho = 6.5^{+2.0}_{-2.7} \; \mathrm{g cm}^{-3}$), Kepler-20c \citep[$\rho = 2.91^{+0.85}_{-1.08} \; \mathrm{g cm}^{-3}$;][]{2012ApJ...749...15G}, and Kepler-68b \citep[$\rho = 3.32^{+0.86}_{-0.98} \; \mathrm{g cm}^{-3}$;][]{2013ApJ...766...40G}, the compositions are still uncertain, although densities greater than $5 \; \rm{g cm}^{-3}$ are consistent with rocky and iron-nickel composition \citep{2012ApJ...749...15G,2014PNAS..11112655M,2014ApJS..210...20M}.  The upper limits on the masses for all the small planets in our sample except for Kepler-20c, Kepler-68b, and Kepler-100c allow for densities greater than $5 \; \rm{g cm}^{-3}$. Moreover, \citet{2014PNAS..11112655M} show that for the known exoplanets with $2\sigma$ mass determinations (33 planets at the time of this writing), those with radii less than $2 \; R_{\earth}$ have densities $\gtrsim 5 \; \rm{g cm}^{-3}$. More recently, \citet{2015ApJ...801...41R} and \citet{2015ApJ...800..135D} find that the transition from rocky to primarily non-rocky planets actually occurs at the smaller radius of $1.6 \; R_{\earth}$. Each of the stars in our sample, save Kepler-22, has at least one planet with a radius less than or approximately equal to $1.6 \; R_{\earth}$.  It is therefore reasonable to assume that with the exception of Kepler-22, each star in our sample is host to at least one planet with a rocky or primarily rocky composition, and it provides a meaningful test of the \tc\ slope-rocky planet correlation hypothesis.

We quantify the relationship between the stellar abundances and \tc\ by measuring the slope of a standard least-squares fit to the refractory element ($T_{\mathrm{c}} > 900$ K) abundances relative to Fe and the 50\% condensation temperatures from \citet{2003ApJ...591.1220L}. Both unweighted fits and fits weighted by the inverse square of the total uncertainties in the [x/Fe] abundances have been made. The slopes and their $1\sigma$ uncertainties are provided in Table \ref{tab:slopes}, and the weighted fits are shown in Figure \ref{fig:tc}. The difference in the slopes from the unweighted and weighted fits are statistically insignificant. Initial inspection of the fits to the data revealed that the Mn abundance falls below the fit to the [x/Fe]-\tc\ relation for each star, by more than the $1\sigma$ total uncertainty in the Mn abundances for three of the stars, despite the fact that all of the Mn abundances are in agreement within uncertainties with the general Galactic population (see Figure \ref{fig:gal}).  Because the \tc\ of Mn is 1158 K and falls near the cool end of the \tc\ range, these low abundances could be biasing the slopes of the fits toward more positive values. We therefore have made unweighted and weighted fits to the data with Mn removed from the list of elements; the slopes of these fits are also included in Table \ref{tab:slopes}. The difference in the slopes from these unweighted and weighted fits are once again found to be statistically insignificant. Moreover, all four slope measurements for each star are consistent within the $1\sigma$ uncertainties.

The slopes for the five MS dwarfs in our sample are all positive at greater than the $2\sigma$ level, whereas the slopes for the two subgiants are negative at greater than the $1.5\sigma$ level, except for the unweighted and weighted slopes for Kepler-21 with the Mn abundance included in the calculation. In these two cases, the slopes are statistically consistent with positive values. We remind the reader that a positive slope in the [x/Fe]-\tc\ plane signifies an overabundance of refractory elements relative to the Sun, and in the interpretation of \citet{2009ApJ...704L..66M} and \citet{2010A&A...521A..33R}, suggests that the star is \textbf{not} a rocky planet host. On the other hand, a flat or negative slope signifies that the star has a similar refractory element distribution as the Sun or is deficient in refractories relative to the Sun, and thus the star would be a candidate for hosting rocky planets. Therefore, the slopes of the five MS dwarf stars are not consistent with the \tc\ slope-rocky planet correlation hypothesis. While the slopes for the two subgiants would appear to be consistent with the hypothesis, these stars' changing internal structure, in particular their deepening convective envelopes, have certainly mixed internal material to their surfaces, potentially changing their photospheric compositions. The \tc-planet signature is expected to be sensitive to the depth of the convective envelope during the planet formation process \citep[e.g.,][]{2010ApJ...724...92C}, so the increase in the depth of the convection zone of an evolving star could alter the signature. We therefore omit the two subgiants, Kepler-21 and Kepler-100, from the remaining discussion in this subsection.

Much of the preceding work investigating the possible correlation between \tc-abundance trends and rocky planets have focused on solar twins and analogs, and late F-type dwarfs. The former because of the similar structure to the Sun, and the latter because of the thinner convective envelopes compared to the Sun. In each case, it is thought that a planet formation signature will persist in the photospheres of these stars \citep[e.g.,][]{2014A&A...561A...7R}, as opposed to less massive stars with larger convective envelopes, in which any signature would be lost to convective mixing on a relatively short timescale. Two of the stars in our sample have convection zones that are estimated to be approximately equal to or smaller than that of the Sun, as discussed below.

The mass of Kepler-68 has been measured to be $1.079 \pm 0.051 \; \mathrm{M}_{\earth}$ \citep{2013ApJ...766...40G}, and thus this star would be expected to have a convective envelope that is at least similar to or thinner than that of the Sun. It is host to an Earth-size planet, a sub-Neptune, and Jupiter-type giant planet. The mass and density of the Earth-size planet, Kepler-68c, have yet to be firmly determined, but \citet{2014ApJS..210...20M} calculate upper limits of $M_{\mathrm{c}} \leq 2.18 \pm 3.5 \; M_{\earth}$ and $\rho_{\mathrm{c}} \leq 10.77 \pm 17.29 \; \mathrm{g \, cm}^{-3}$ based on four years of RV observations. Given these upper limits and its size, it is reasonable to assume that Kepler-68c is a rocky planet. The mass and density of the sub-Neptune planet, Kepler-68b, have been determined, and at $M_{\mathrm{b}} = 5.97 \pm 1.7 \; M_{\earth}$ and $\rho_{\mathrm{b}} = 2.60 \pm 0.74 \; \mathrm{g \, cm}^{-3}$, the planet's density falls in between those of gas giant and rocky planets. \citet{2014ApJ...792....1L} suggest that planets like Kepler-68b represent a transition between rocky planets and gas giant planets, and may be composed of large quantities of (ice/liquid) water while lacking a large H/He gas envelope. The planetary models of \citet{2014ApJ...792....1L} and \citet{2014ApJ...787..173H} suggest that if Kepler-68b has a H/He envelope at all, it makes up less than $2\%$ of the planet mass. Using the model of \citet{2007ApJ...659.1661F}, the composition of Kepler-68b is predicted to be close to $75\%$ water ice and $25\%$ rock. Assuming a $25\%$ rocky composition for Kepler-68b and a rocky composition for Kepler-68c, a total of $3.7 \; M_{\earth}$ of rocky material is sequestered in these two planets. We do not include the giant planet Kepler-68d in this calculation, because it is unclear if large, gas giant planets contribute to the putative depletion of refractory elements in the host star \citep{2011ApJ...732...55S,2014A&A...561A...7R}. The amount of refractory material contained in Kepler-68b and Kepler-68c is more than what has been suggested to be missing in the photosphere of the Sun, so if rocky planet formation depletes the refractory elements in host stars, we would expect to observe such a depletion in Kepler-68. 

We have tested this expectation by calculating the changes in the photospheric abundances of Kepler-68 due to the extraction of $3.7 \; M_{\earth}$ of material from its convection zone. Using the accretion prescription of \citet{2014ApJ...787...98M}, we calculated the abundance changes in the star assuming the accretion of $3.7 \; M_{\earth}$ of material and then subtracted the changes from the observed abundances to estimate the effects on the slope of the \tc-abundance relation if this amount of material had indeed been sequestered in the planets during the formation of Kepler-68b and Kepler-68c. In addition to the mass of material, the accretion model inputs include the composition of the material, the mass of the star, and the mass of the convection zone. We adopted two different compositions of the extracted material, one that matches the Earth's composition (Earth model) based on \citet{mcdonough2001} and, following \citet{2010ApJ...724...92C}, one composed of 50\% Earth's composition and 50\% carbonaceous (CM) chondrite material (50/50 model), with the CM chondrite compositions taken from \citet{1988RSPTA.325..535W}. Using the mass of Kepler-68 given above, the mass of its convection zone was estimated from Figure 1 of \citet{2001ApJ...556L..59P} using our derived \teff. The results of the modeled abundance changes are shown in Figure \ref{fig:tc_68}. Both the Earth model and the 50/50 model predict abundance changes that result in slopes of the fits to the \tc-abundance data that are consistent with zero. According to the \tc\ slope-rocky planet correlation hypothesis, a zero slope would be indicative of rocky planet formation. This result reinforces our assertion that if rocky planet formation does in fact deplete the refractory elements in host stars, we would expect to detect the signature in Kepler-68.

The second star with a convective envelope that is estimated to be roughly equal to the Sun's is Kepler-130. Density ($\rho = 0.927 \pm 0.053 \; \mathrm{g \, cm}^{-3}$) and radius ($R = 1.127 \pm 0.033 \; R_{\odot}$) estimates for Kepler-130 based on posterior distributions of stellar evolution models are presented by \citet{2014ApJ...784...45R}; however, these estimates are based on the stellar parameters of \citet{2013ApJ...767..127H}, which are about 75 K and 0.11 dex in \teff\ and $\log g$, respectively, lower than ours (although, as mentioned above, these are in agreement with ours within the combined uncertainty). Using their radius, density, and $\log g$, as well as our derived $\log g$, we calculate the mass of Kepler-130 to range from 0.932 $M_{\odot}$ to 1.19 $M_{\odot}$. Thus, assuming a convective envelope on par with that of the Sun's is not unreasonable. Kepler-130 is host to an Earth-size planet, a super-Earth, and a sub-Neptune; unfortunately, definitive masses have not been determined for any of them. However, the small sizes of the planets, particularly Kepler-130b ($R = 1.02 \pm 0.04 \; R_{\earth}$) and Kepler-130d ($R = 1.64 \pm 0.16 \; R_{\earth}$), are suggestive of rocky compositions. Using the empirical mass-radius relation of \citet{2014ApJ...783L...6W}, we estimate the masses of the three planets to be $M_{\mathrm{b}} = 1.1 \; M_{\earth}$, $M_{\mathrm{c}} = 7.0 \; M_{\earth}$, and $M_{\mathrm{d}} = 4.3 \; M_{\earth}$ for Kepler-130b, Kepler-130c, and Kepler-130d, respectively. We then use the model of \citet{2007ApJ...659.1661F} to predict the compositions of the planets and find an Earth-like composition for Kepler-130b, a pure ice composition for Kepler-130c, and a pure rock composition for Kepler-130d. Considering Kepler-130b and Kepler-130d only, as much as 5.4 $M_{\earth}$ of refractory material could be sequestered in these planets. Adopting a stellar mass of $1.061 \; \mathrm{M}_{\earth}$ (average of the range given above) and estimating a convection zone mass using \citet{2001ApJ...556L..59P}, we use the Earth and 50/50 composition models described above to determine the effect on the slope of the \tc-abundance relation if this amount of material had been sequestered in Kepler-130b and Kepler-130d. The results are shown in Figure \ref{fig:tc_130}. As found for Kepler-68, both the Earth model and 50/50 model predict abundance changes that result in \tc-abundance slopes that are consistent with zero, suggesting that if this star's refractory elements were depleted due to small planet formation, we should be able to detect it.

One more system, Kepler-20, also warrants additional scrutiny, because masses and densities have been well determined for two of its planets \citep{2012ApJ...749...15G}. The density of Kepler-20b is $\rho_{\mathrm{b}} = 6.5^{+2.0}_{-2.7} \; \mathrm{g \, cm}^{-3}$, placing this planet squarely in the rocky planet category, and that of Kepler-20c is $\rho_{\mathrm{c}} = 2.91^{+0.85}_{-1.08} \; \mathrm{g \, cm}^{-3}$, placing it in the transitional water-world regime. Two more planets in this system, Kepler-20e and Kepler-20f, have radii near that of Earth, so it is reasonable to assume they too are rocky planets. The fifth planet in the system, Kepler-20d, has a radius equal to that of Kepler-20c within uncertainties and is probably of the water-world type planet, as well. Once again we use the empirical mass-radius relation of \citet{2014ApJ...783L...6W} to estimate the masses of Kepler-20d ($M_{\mathrm{d}} = 6.9 \; M_{\earth}$), Kepler-20e ($M_{\mathrm{e}} = 0.64 \; M_{\earth}$), and Kepler-20f ($M_{\mathrm{f}} = 1.2 \; M_{\earth}$) and the model of \citet{2007ApJ...659.1661F} to predict the compositions of all five planets. Compositions of the planets are found to be pure rocky for Kepler-20b, $75\%$ water ice and $25\%$ rocky for Kepler-20c, pure ice for Kepler-20d, and Earth-like compositions ($\sim25\%$ rocky and $\sim75\%$ iron) for both Kepler-20e and Kepler-20f. If we include only the lower limits of the masses of Kepler-20b and Kepler-20c, a total of 9.6 $M_{\earth}$ of refractory material is contained in these two planets, whereas if we use the upper limits of the masses for these two planets and include the estimated masses of Kepler-20e and Kepler-20f, as much as 17.5 $M_{\earth}$ of refractory material could be sequestered in these planets. The mass and radius of the host star have been estimated at $M = 0.912 \pm 0.034 \; M_{\odot}$ and $R = 0.944^{+0.060}_{-0.095} \; R_{\odot}$ \citep{2012ApJ...749...15G,2012Natur.482..195F}, and thus it has a larger convective envelope than the Sun. Nonetheless, using \citet{2001ApJ...556L..59P} to estimate the mass of the star's convection zone, both the Earth and 50/50 composition models predict \tc-abundance slopes that are negative or consistent with zero if either 9.6 $M_{\earth}$ or 17.5 $M_{\earth}$ of refractory material has been sequestered in the Kepler-20 planets (Figure \ref{fig:tc_20}). Once again we find that the putative rocky planet signature, if real, should be detectable in this star.

\section{SUMMARY}
\label{s:summary}
Stellar parameters and detailed abundances of 19 elements have been derived for seven stars known to host small planets discovered by NASA's \kep\ Mission. We have analyzed high-quality spectra obtained with the 10-m Keck telescope and HIRES echelle spectrometer as part of the \kep\ Follow-up Observing Program and independent observations. Previously derived stellar parameters (\teff\ and $\log g$) and [Fe/H] abundances by numerous other groups using automated and semi-automated routines (SME, SPC, ARES, VWA, etc.) with high-resolution spectroscopy are, in general, in good agreement with our derived values. In seven out of about 36 comparisons ($\sim 19\%$), the stellar parameters derived using automated or semi-automated routines are different than ours derived ``by hand'' by more than the combined uncertainties. However, in each case, the derived [Fe/H] abundances are in agreement within the uncertainties. Differences between stellar parameters and [Fe/H] abundances derived for two stars using low-resolution spectroscopic analysis and our derived values are more disparate, and while considered to be an improvement over the photometrically derived values in the \kep\ Input Catalog, results based on low-resolution spectroscopy should be used with caution. Our surface gravities are in good agreement with asteroseismic derived values for four of the six stars but are larger by about the $1.7\sigma$ for the other two. Nonetheless, the derived \teff\ and abundances for these two stars remain robust, because they do no correlate with changes in $\log g$ and due to the small sensitivities to the $\log g$ parameter.

The metallicities of the seven stars range from [Fe/H] $= -0.30$ to $+0.15$, further demonstrating that small planets form around stars with both low and high metallicities \citep{2012Natur.486..375B,2015ApJ...808..187B}. The abundances of 15 elements are compared to those of a Galactic disk population that includes stars with and without known planets. The abundances of the \kep\ planet host stars fall along the Galactic trends, suggesting that stars with small planets have compositions that are typical of the Galactic disk. Moreover, comparing the abundance distributions of the \kep\ planet host stars to a sample of stars with no detected planets, a sample of Neptune- and super-Earth-size planet hosts, and a sample of Jupiter-type giant planet hosts, reveals that the former three-- \kep\ host stars, stars without known planets, and the Neptune and super-Earth-size planet hosts-- may have similar compositions, whereas the \kep\ host stars appear to have compositions that are less enhanced than those of Jupiter-type planet hosts. These results suggest that the formation of small planets does not require exceptional host-star or protoplanetary disk compositions, and that small planets may indeed be ubiquitous in the Galaxy. Our small sample size limits the statistical significance of our results, so a larger sample of stars with small planets subjected to detailed abundance analyses is needed to verify these preliminary findings.

The abundances of the refractory elements ([x/Fe]) as a function of \tc\ have been examined. Slopes of least-square fits to the data are found to be positive at greater than the $2\sigma$ level for the five MS dwarfs in our sample. This is contrary to expectations of the hypothesis that rocky planet formation sequesters refractory element material from host stars \citep{2009ApJ...704L..66M,2009A&A...508L..17R}. Using known planet masses when available and when not, masses determined by an empirical mass-radius relation \citep{2014ApJ...783L...6W}, we make use of planet composition models \citep{2007ApJ...659.1661F} to estimate the amount of refractory material contained in the rocky planets of three systems- Kepler-68, Kepler-130, and Kepler-20. In each case, we estimate that there is enough refractory material in these planets to produce an observable flat or negative trend in the [x/Fe]-\tc\ data if in fact small planet formation imprints such a signature on their host star. We do not see this signature for these stars.

The simplest interpretation of the positive \tc\ slopes we measure for these five dwarf stars known to host small planets is that there is no direct connection between the presence of small planets and depleted refractory elements in the photospheres of their host stars. However, a definitive conclusion on the possible connection between the presence of rocky planets and the detailed composition of a stellar atmosphere cannot be made based solely on our modest sample size, but our results do indicate that there is not a simple correlation between the two.  The first suggestion that rocky planet formation may affect the refractory element abundances of the host star resulted from a fine analysis of the Sun \citep{2009ApJ...704L..66M}, the planetary architecture of which is unlike any of the stars in our sample.  Our stars all have planets with semimajor axes smaller than that of Mercury, with the exception of the sub-Neptune Kepler-22b \citep[$a = 0.85$ AU][]{2012ApJ...745..120B} and the giant planet Kepler-68d \citep[$a = 1.4$ AU][]{2013ApJ...766...40G}, so the architecture of a planetary system may play an important role. Also, factors such as stellar age, stellar structure, Galactic chemical evolution, and stellar birthplace in the Galaxy could all affect the composition of a star \citep{2014A&A...561A...7R,2014A&A...564L..15A,2015A&A...579A..52N}. The planet formation process is undoubtedly complex, and determining if small planet formation in particular alters the compositions of the host stars has revealed itself to be quite challenging. Detailed abundance analyses of stars known to host small planets offer the best way to resolve the current discordance in the field, and with the \kep\ mission successfully identifying numerous small planets, such analyses are now possible.

Finally, the seeming lack of dependence of small planet formation on metallicity and the concomitant propensity of giant planets to form around metal-rich stars are characteristic of core accretion models of planet formation \citep[e.g.,][]{2004ApJ...616..567I}, as demonstrated by planet population syntheses \citep[e.g.,][]{2012A&A...541A..97M}. The core accretion paradigm is the most widely accepted model of planet formation, but many details still need to be worked out \citep[e.g.,][]{2013ApJ...775...42I,2013A&A...549A..44F} if a single model is to explain the observed properties of the known planetary systems. Furthermore, the competing paradigm of planet formation, gravitational instability \citep[e.g.,][]{1997Sci...276.1836B}, may yet play a role in any unified theory of planet formation \citep[e.g.,][]{2013MNRAS.432.3168F}.  The continued characterization of planetary systems, including the detailed compositions of host stars, will be necessary to guide future efforts in fashioning such a theory.

\acknowledgements
We graciously thank F. Valdes at NOAO for his help with the processing of the Keck/HIRES data, as well as G. Marcy and A. Howard for their role in obtaining the FOP/Keck spectra. S.C.S. acknowledges support provided by grant NNX12AD19G from the National Aeronautics and Space Administration as part of the Kepler Participating Scientist Program. Some data presented herein were obtained at the W. M. Keck Observatory from telescope time allocated to the National Aeronautics and Space Administration through the agency's scientific partnership with the California Institute of Technology and the University of California. The Observatory was made possible by the generous financial support of the W. M. Keck Foundation. Additional data were obtained at Kitt Peak National Observatory, National Optical Astronomy Observatory, which is operated by the Association of Universities for Research in Astronomy (AURA) under cooperative agreement with the National Science Foundation.  This research has made use of the NASA Exoplanet Archive, which is operated by the California Institute of Technology, under contract with the National Aeronautics and Space Administration under the Exoplanet Exploration Program.

{\it Facilities:} \facility{Keck:I (HIRES)} \facility{KPNO:4.0m (Echelle)}

\appendix
\section{Appendix}
\label{s:append}
Three different group members measured EWs for three separate sets of stars in our sample, and in order to ensure as much consistency as possible in the final relative abundances, each person also measured EWs in the solar spectrum appropriate for each star. The measured EWs and resulting absolute abundances for each line of all elements for each star are provided in Table \ref{tab:lines}, and the three sets of solar EWs and absolute abundances are provided in Table \ref{tab:slines}.

\newpage

\begin{figure}
\plotone{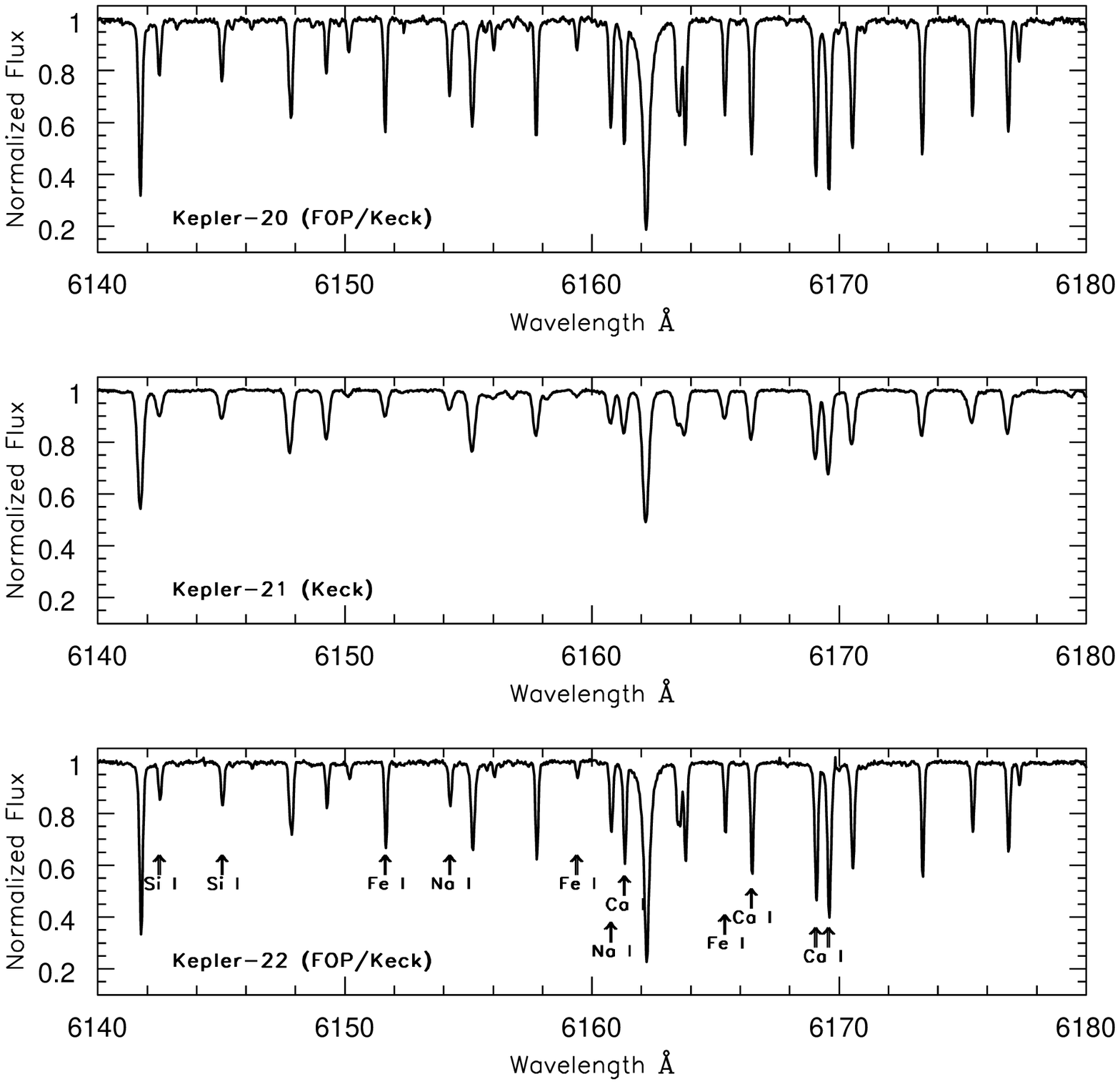}
\caption{Sample Keck/HIRES spectra of some of the \kep\ stars. Lines for which EWs were measured are marked.\label{fig:spec}}
\end{figure}

\begin{figure}
\plotone{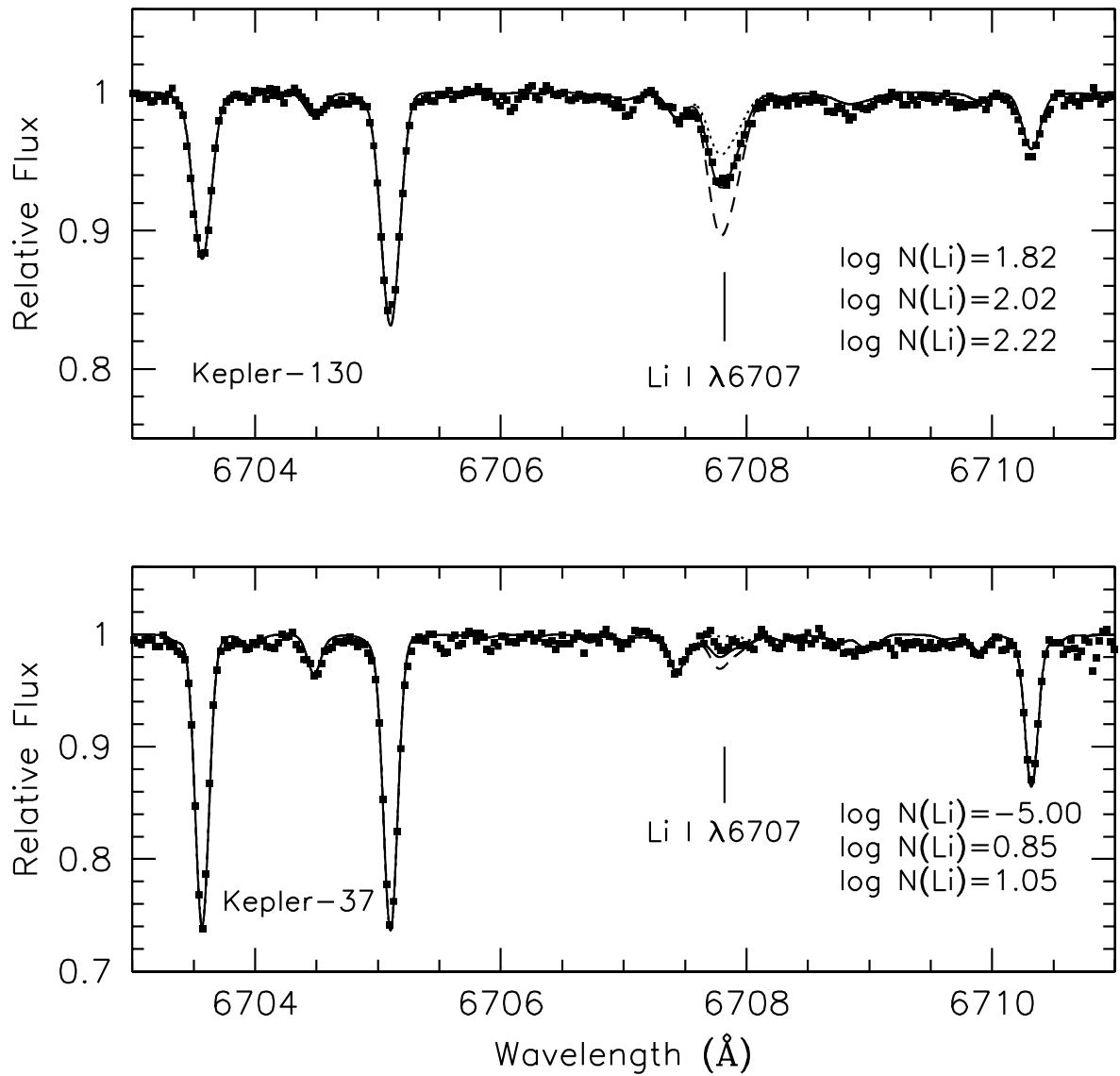}
\caption{Observed spectra (points) and synthetic fits (lines) of the $\lambda 6707$ \ion{Li}{1} region of Kepler-37 (bottom panel) and Kepler-130 (top panel). Syntheses are shown for three input Li abundances for each star. For Kepler-130, the best fit abundance and $\pm 0.20$ dex around the best fit abundance are shown. In the case of Kepler-37, for which a definitive Li measurement was not possible, syntheses for no Li, an upper limit on the Li abundance, and 0.20 dex larger than the upper limit are provided. \label{fig:li}}
\end{figure}

\begin{figure}
\plotone{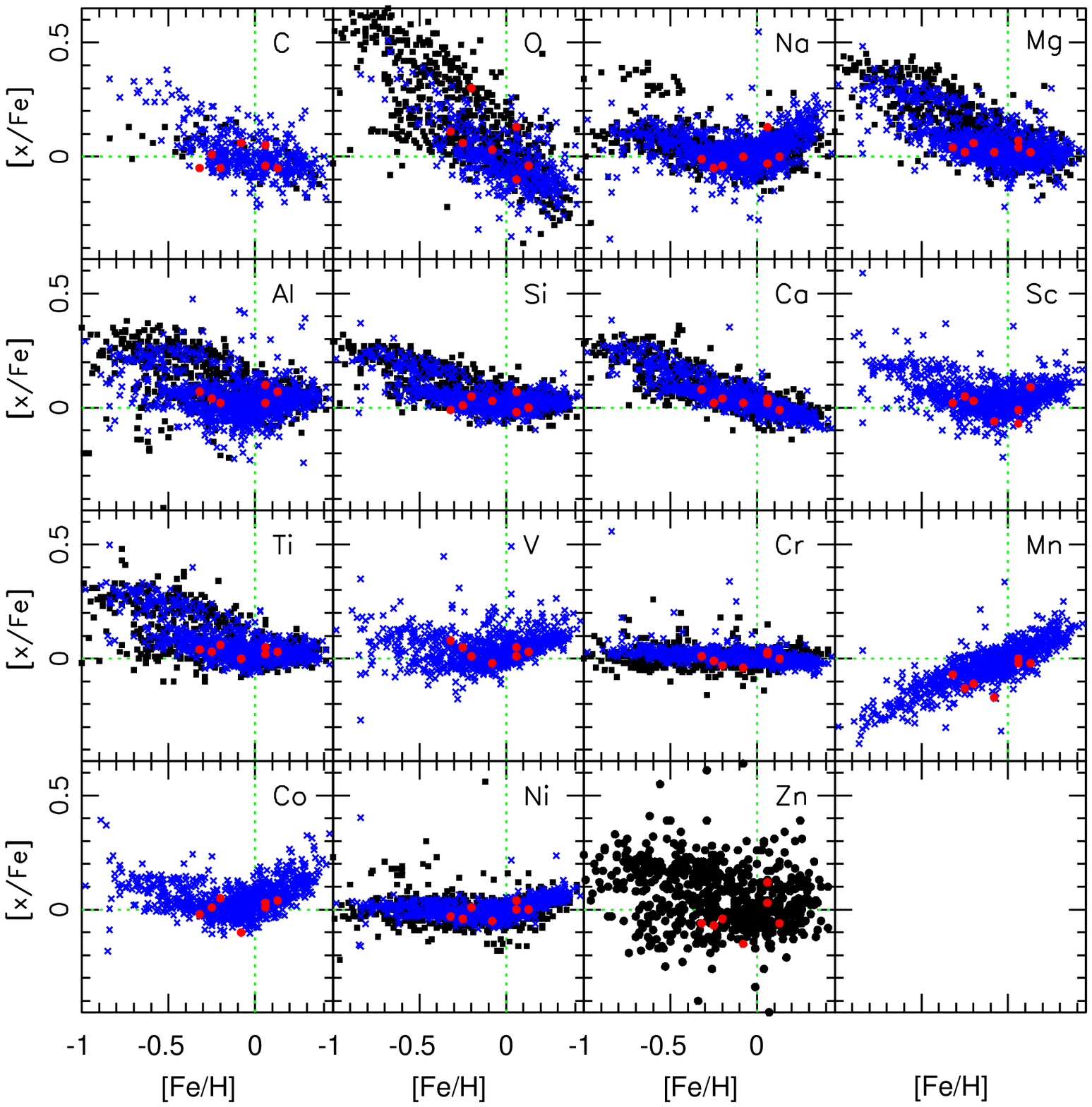}
\caption{[x/Fe] ratios as a function of [Fe/H].  The red circles are the \kep\ stars. For C and O, the blue crosses are abundances taken from \citet{2010ApJ...725.2349D}; black squares are from \citet{2006MNRAS.367.1181B} and the large sample of \citet{2014A&A...562A..71B}, repspectively. For the remaining elements, blue crosses are from \citet{2012A&A...545A..32A}, and black squares are from \citet{2014A&A...562A..71B}. Solar values are indicated by the green dotted lines.\label{fig:gal}}
\end{figure}

\begin{figure}
\plotone{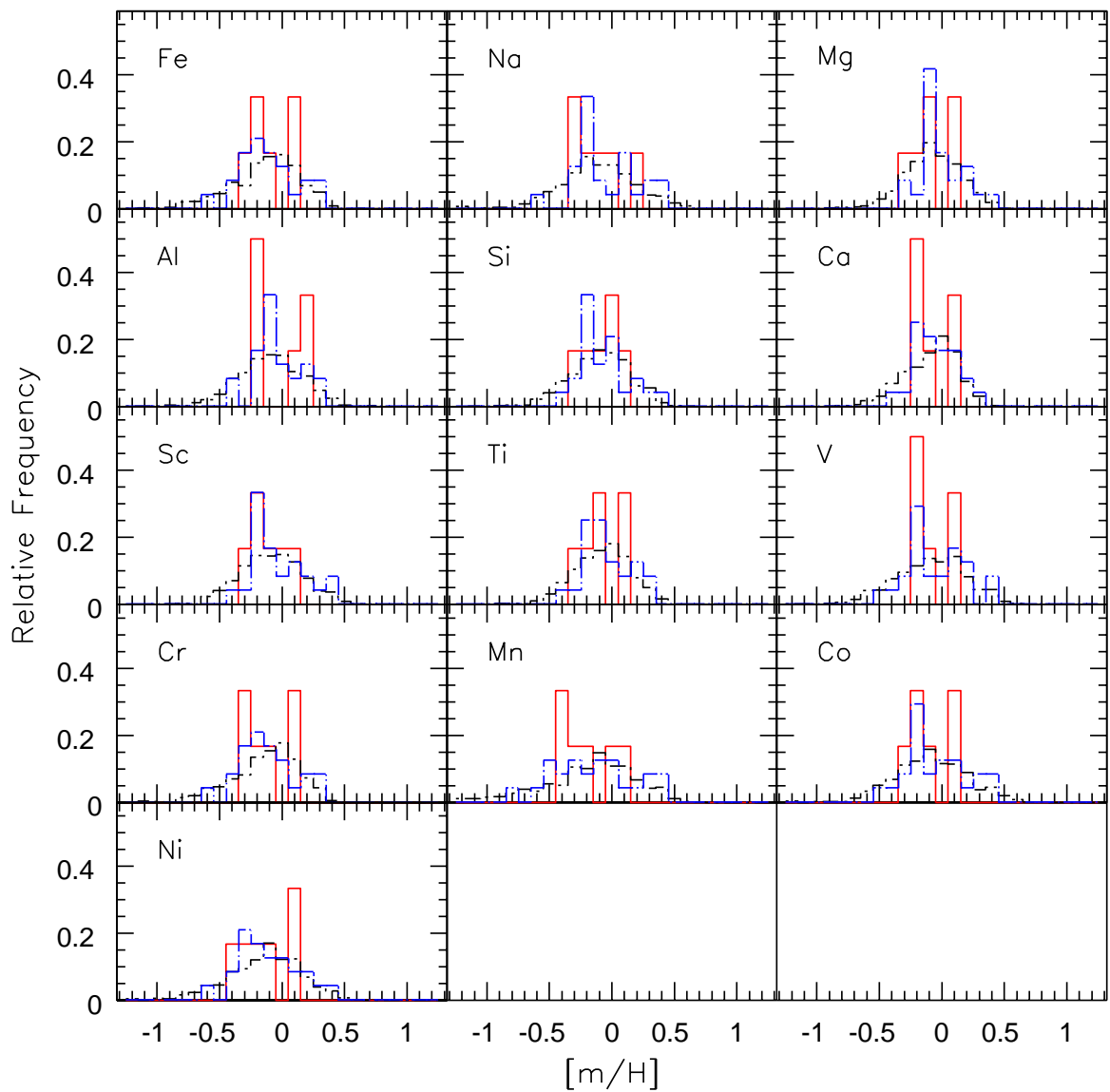}
\caption{Comparison between abundances of the Kepler host stars (red solid lines) and those of stars without known planets (black dot-dash lines) and stars with Neptune- and super-Earth-size planets (blue dot-dash lines). The data for the stars without known planets and with Neptune- and super-Earth-size planets are from \citet{2012A&A...545A..32A}. There is significant overlap of all three histograms for each element.\label{fig:histo_neptunes}}
\end{figure}

\begin{figure}
\plotone{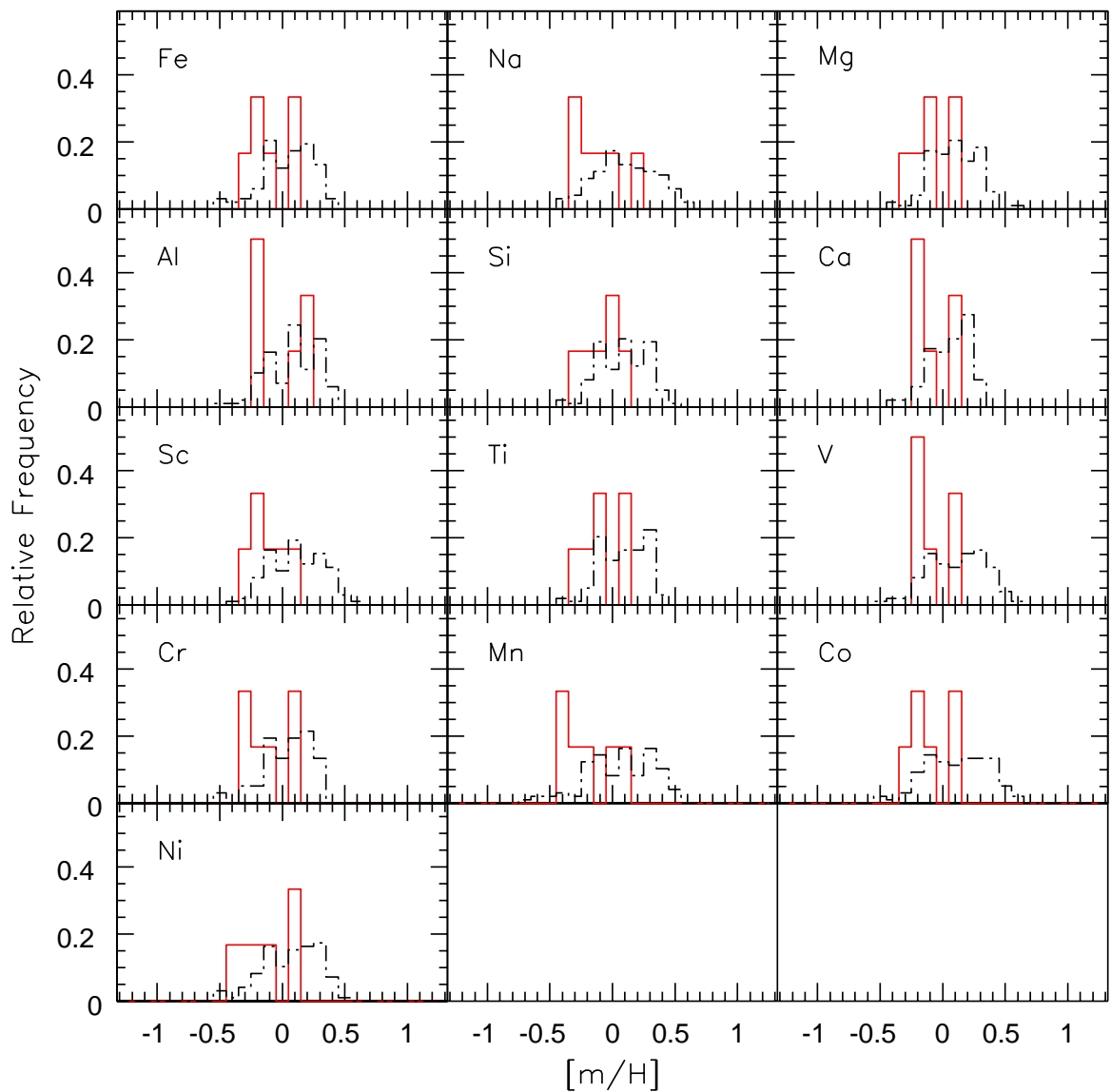}
\caption{Comparison between abundances of the Kepler host stars (red solid lines) and those of stars with Jupiter-type giant planets (black dot-dash lines). The data for the stars with Jupiter-type giant planets are from \citet{2012A&A...545A..32A}. Despite the small sample of Kepler host stars, there appears to be an offset between the two samples, with the Jupiter-type planet hosts shifted to higher abundances.\label{fig:histo_jupiters}}
\end{figure}

\begin{figure}
\plotone{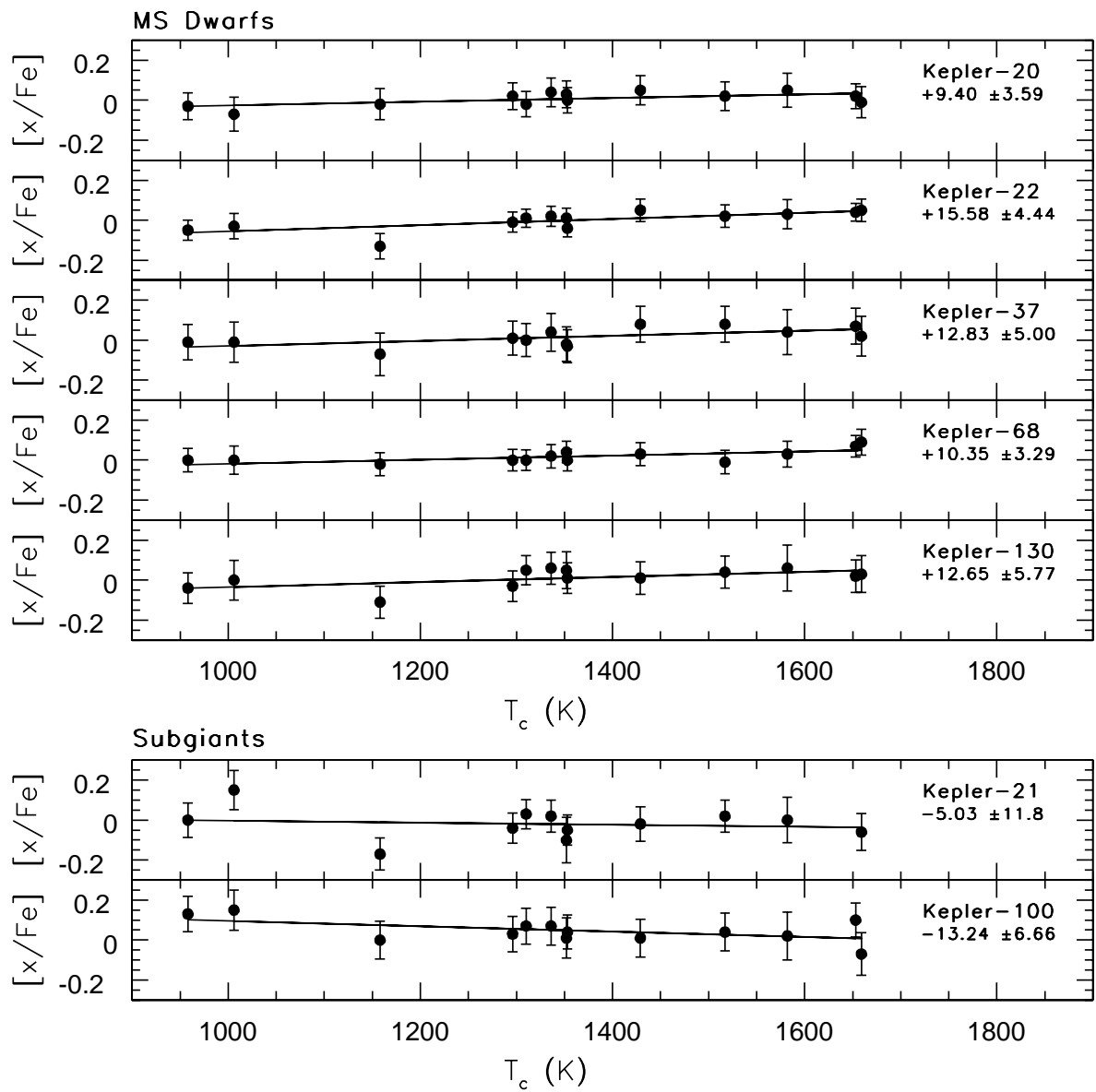}
\caption{Abundances versus condensation temperature of the elements. The solid line is a linear least-squares fit to the data weighted by the inverse square of the total uncertainty in each abundance. The MS dwarfs are displayed in the top group, and the subgiants are shown in the bottom group. The slope and uncertainty of the fit are also provided, in units of $\times 10^{-5} \; \mathrm{dex} \; \mathrm{K}^{-1}$ \label{fig:tc}}
\end{figure}

\begin{figure}
\plotone{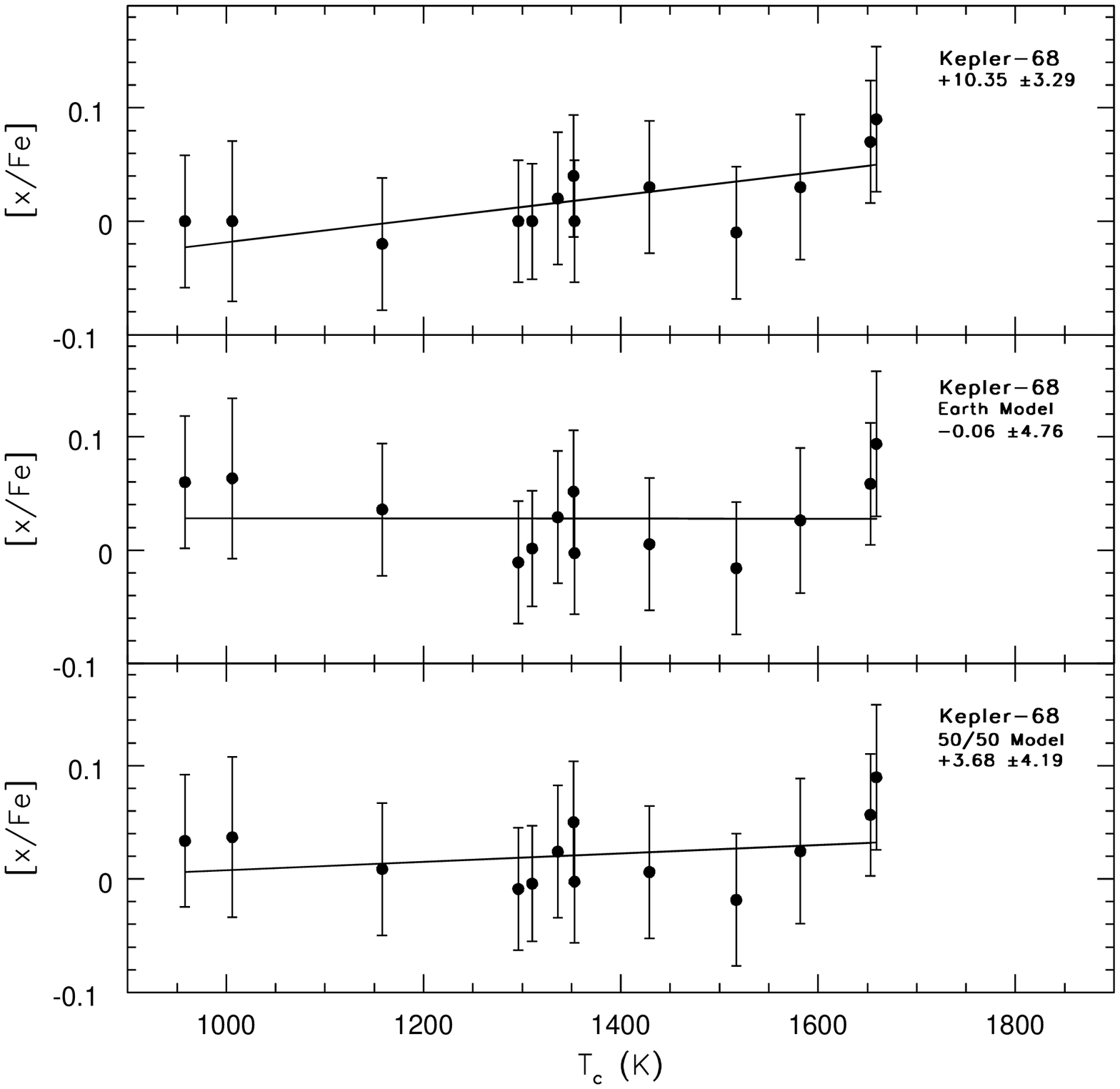}
\caption{Abundances versus condensation temperature of the elements for Kepler-68. The top panel shows the derived abundances, which are identical to those in Figure \ref{fig:tc}. The middle panel shows the modeled abundances after removing $3.7 \; M_{\earth}$ of material with Earth-like composition, and the bottom panel shows the modeled abundances after removing the same amount of material composed of 50\% Earth-like composition and 50\% CM chondrite composition, as described in the text. The solid lines are the linear least-squares fits to the data weighted by the inverse square of the total uncertainty in each abundance. The slope and uncertainty of the fit are also provided, in units of $\times 10^{-5} \; \mathrm{dex} \; \mathrm{K}^{-1}$. For both cases of the modeled abundances, the resulting slopes of the least-squares fits are negative or consistent with zero, which would be in concordance with the \tc\ slope-rocky planet correlation hypothesis, if real. \label{fig:tc_68}}
\end{figure}

\begin{figure}
\plotone{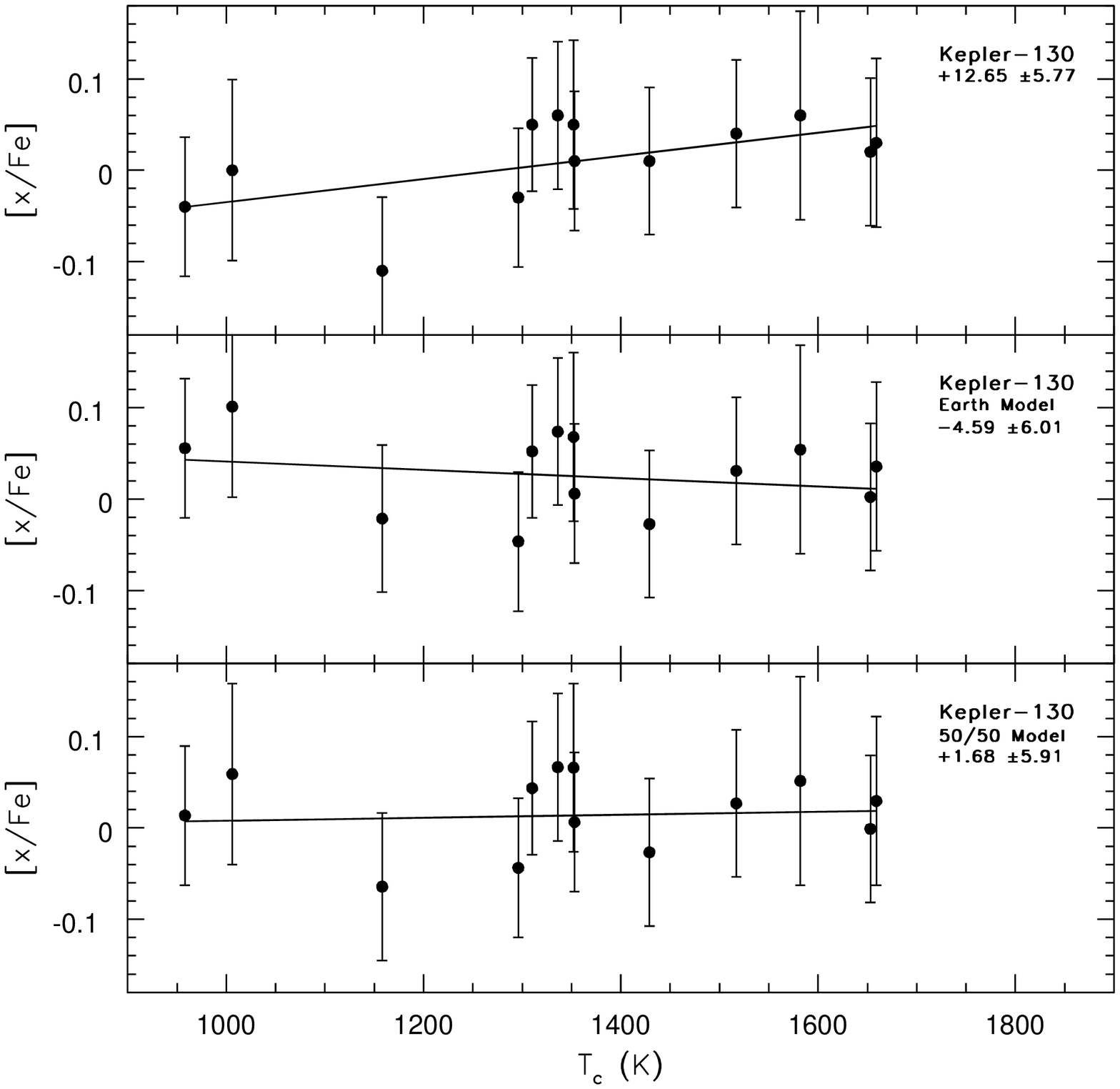}
\caption{Abundances versus condensation temperature of the elements for Kepler-130. The top panel shows the derived abundances, which are identical to those in Figure \ref{fig:tc}. The middle panel shows the modeled abundances after removing $5.4 \; M_{\earth}$ of material with Earth-like composition, and the bottom panel shows the modeled abundances after removing the same amount of material composed of 50\% Earth-like composition and 50\% CM chondrite composition, as described in the text. The solid lines are the linear least-squares fits to the data weighted by the inverse square of the total uncertainty in each abundance. The slope and uncertainty of the fit are also provided, in units of $\times 10^{-5} \; \mathrm{dex} \; \mathrm{K}^{-1}$. For both cases of the modeled abundances, the resulting slopes of the least-squares fits are negative or consistent with zero, which would be in concordance with the \tc\ slope-rocky planet correlation hypothesis, if real. \label{fig:tc_130}}
\end{figure}

\begin{figure}
\plottwo{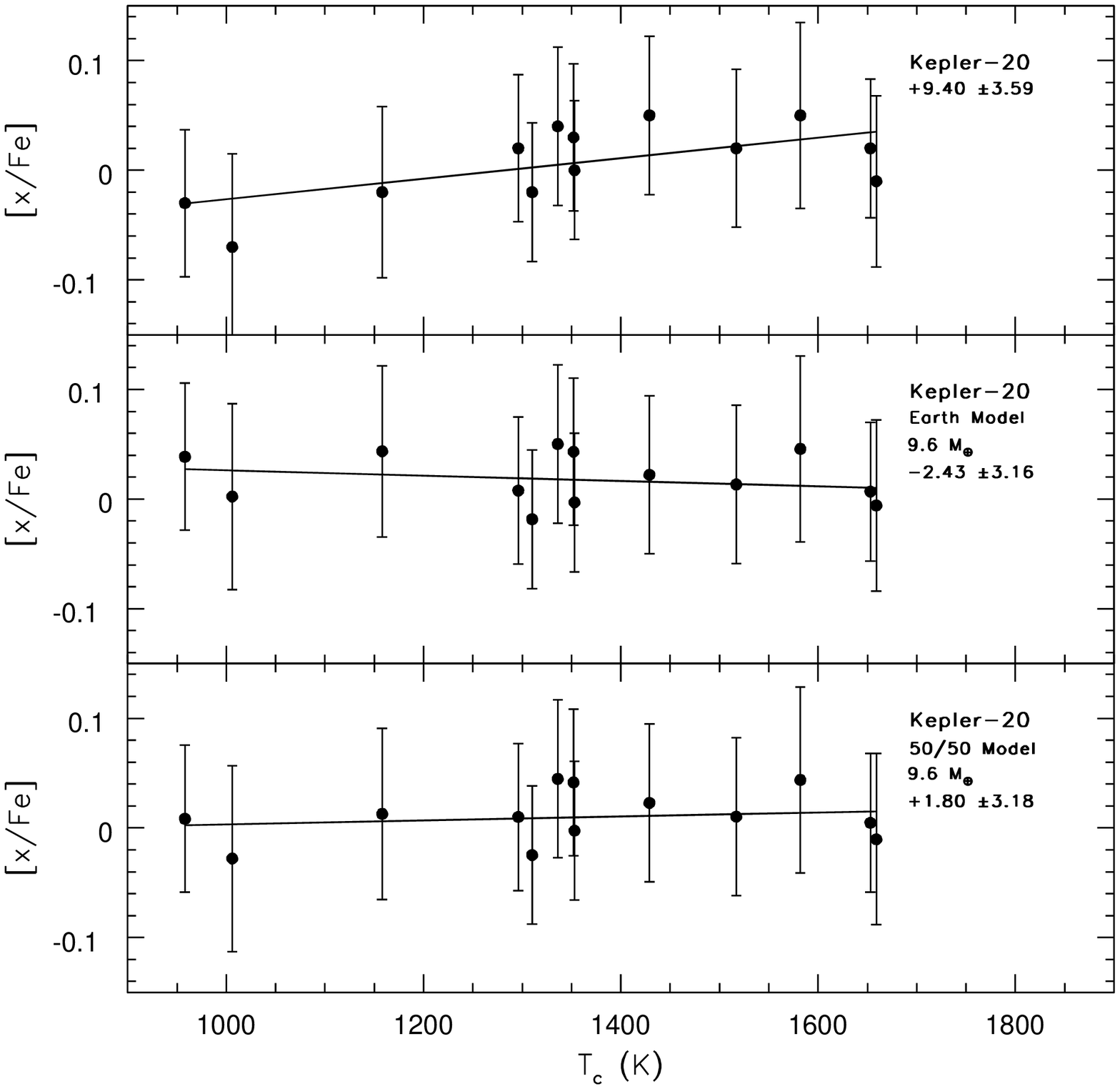}{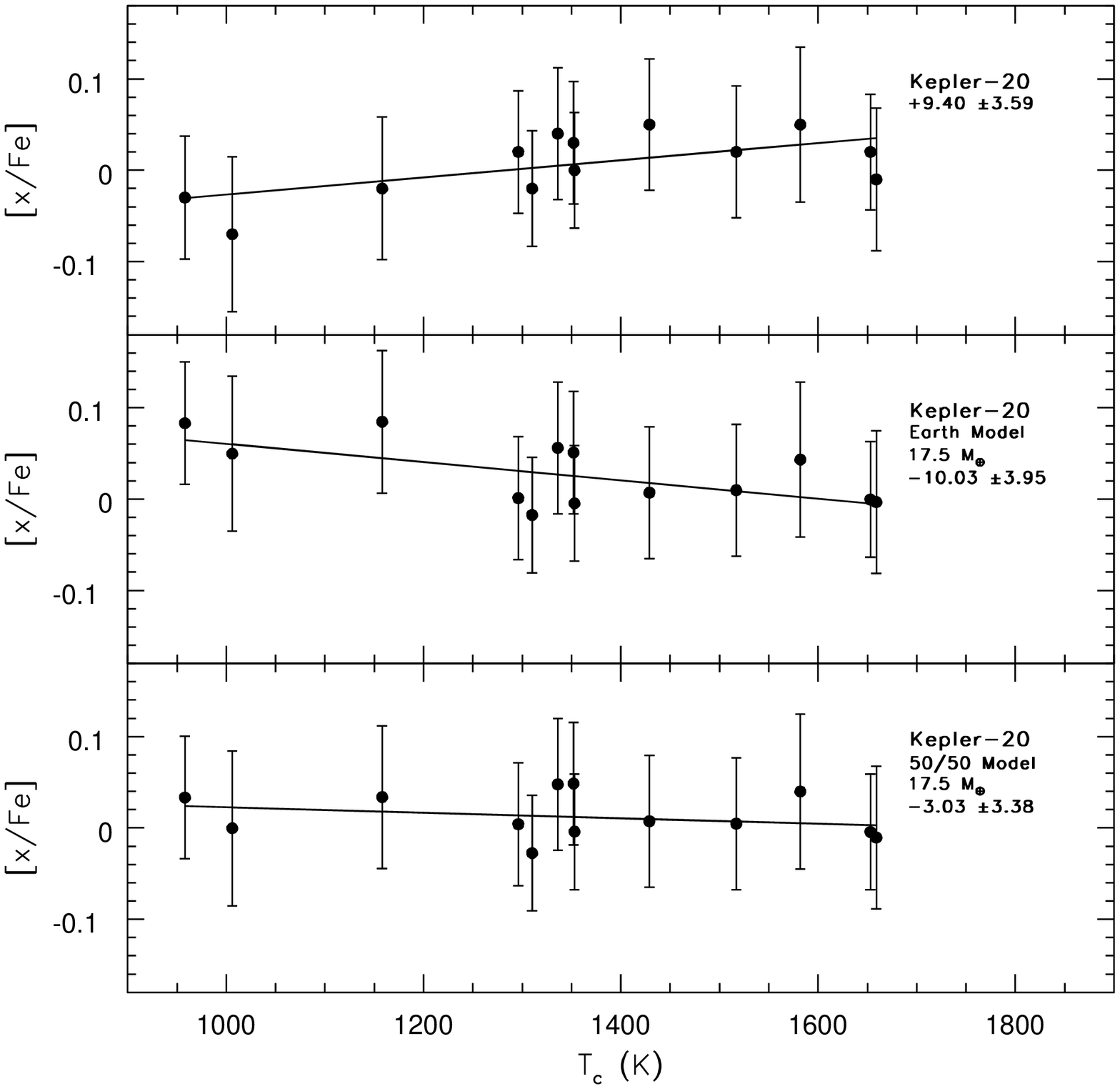}
\caption{Abundances versus condensation temperature of the elements for Kepler-20. The top panels show the derived abundances, which are identical to those in Figure \ref{fig:tc}. The middle panels show the modeled abundances after removing $9.6 \; M_{\earth}$ (left) and $17.5 \; M_{\earth}$ (right) of material with Earth-like composition, and the bottom panels show the modeled abundances after removing the same amount of material composed of 50\% Earth-like composition and 50\% CM chondrite composition, as described in the text. The solid lines are the linear least-squares fits to the data weighted by the inverse square of the total uncertainty in each abundance. The slope and uncertainty of the fit are also provided, in units of $\times 10^{-5} \; \mathrm{dex} \; \mathrm{K}^{-1}$. For all cases of the modeled abundances, the resulting slopes of the least-squares fits are negative or consistent with zero, which would be in concordance with the \tc\ slope-rocky planet correlation hypothesis, if real. \label{fig:tc_20}}
\end{figure}


\begin{deluxetable}{lcrcrcrcccccl}
\tablecolumns{13}
\tablewidth{0pt}
\tablecaption{Observing Log\label{tab:obs}}
\tablehead{
	\colhead{Star}&
	\colhead{}&
	\colhead{KOI}&
	\colhead{}&
	\colhead{KIC}&
	\colhead{}&
	\colhead{$K_{\mathrm{p}}$ \tablenotemark{a}}&
	\colhead{}&
	\colhead{Date}&
	\colhead{}&
	\colhead{S/N \tablenotemark{b}}&
	\colhead{}&
	\colhead{Comment}\\
	\colhead{}&
	\colhead{}&
	\colhead{}&
	\colhead{}&
	\colhead{}&
	\colhead{}&
	\colhead{}&
	\colhead{}&
	\colhead{(UT)}&
	\colhead{}&	
	\colhead{}&
	\colhead{}&
	\colhead{}
	}
\startdata
Kepler-20 && 70 && 6850504 && 12.5 && 2010 Aug 24 && 231 && FOP/Keck \\
Kepler-21 &&975 && 3632418 &&  8.2 && 2011 Jul 12 && 482\tablenotemark{c} && Keck \\
          &&	&&	   &&	   && 2010 Sep 01 && 160 && FOP/Keck \\
          &&	&&	   &&	   && 2011 Apr 19 && 435 && KPNO \\
Kepler-22 && 87 &&10593626 && 11.7 && 2010 Sep 25 && 314\tablenotemark{d} && FOP/Keck \\
          &&	&&	   &&	   && 2011 Sep 05 &&	 && FOP/Keck \\
Kepler-37 &&245 && 8478994 &&  9.7 && 2010 Aug 02 && 258 && FOP/Keck \\
Kepler-68 &&246 &&11295426 && 10.0 && 2010 Aug 02 && 250 && FOP/Keck \\
Kepler-100&& 41 && 6521045 && 11.0 && 2012 Aug 31 && 340 && Keck \\
Kepler-130&&282 && 5088536 && 11.5 && 2012 Aug 31 && 290 && Keck \\
\enddata

\tablenotetext{a}{\citet{2011ApJ...736...19B}}
\tablenotetext{b}{S/N estimates given at 6700 {\AA}}
\tablenotetext{c}{S/N at 6850 {\AA}}
\tablenotetext{d}{S/N of combined FOP/Keck sepctra}

\end{deluxetable}

\begin{deluxetable}{lcccccccrcccrrccl}
\tablecolumns{17}
\tablewidth{0pt}
\rotate
\tablecaption{Stellar Parameters\label{tab:params}}
\tablehead{
	\colhead{Star}&
	\colhead{}&
	\colhead{\teff}&
	\colhead{}&
	\colhead{$\log g$}&
	\colhead{}&
	\colhead{$\xi$}&
	\colhead{}&
	\colhead{[\ion{Fe}{1}/H]}&
	\colhead{$N$}&
	\colhead{$\sigma_{\mu}$}&
	\colhead{}&
	\colhead{[\ion{Fe}{2}/H]}&
	\colhead{$N$}&
	\colhead{$\sigma_{\mu}$}&
	\colhead{}&
	\colhead{Comment}\\
	\colhead{}&
	\colhead{}&
	\colhead{(K)}&
	\colhead{}&
	\colhead{(cgs)}&
	\colhead{}&
	\colhead{(km s$^{-1}$)}&
	\colhead{}&
	\colhead{}&
	\colhead{}&
	\colhead{}&
	\colhead{}&
	\colhead{}&
	\colhead{}&
	\colhead{}&
	\colhead{}&
	\colhead{}
	}
\startdata
Kepler-20 && 5514$\pm 25$ && 4.44$\pm 0.12$ && 1.15$\pm 0.04$ && +0.059 & 61 & 0.003 && +0.062 & 13 & 0.009 && FOP/Keck\\
Kepler-21 && 6177$\pm 42$ && 3.99$\pm 0.15$ && 1.98$\pm 0.15$ && -0.079 & 48 & 0.006 && -0.079 & 13 & 0.011 && Keck\\
          && 6125$\pm 45$ && 3.90$\pm 0.18$ && 1.95$\pm 0.16$ && -0.102 & 46 & 0.006 && -0.096 & 14 & 0.012 && FOP/Keck\\
          && 6240$\pm 45$ && 4.18$\pm 0.14$ && 1.82$\pm 0.20$ && -0.057 & 44 & 0.007 && -0.060 &  5 & 0.017 && KPNO\\
Kepler-22 && 5622$\pm 30$ && 4.57$\pm 0.08$ && 1.47$\pm 0.12$ && -0.251 & 61 & 0.004 && -0.250 & 14 & 0.012 && FOP/Keck\\
Kepler-37 && 5406$\pm 28$ && 4.49$\pm 0.13$ && 1.36$\pm 0.06$ && -0.323 & 58 & 0.004 && -0.321 & 13 & 0.020 && FOP/Keck\\
Kepler-68 && 5887$\pm 28$ && 4.45$\pm 0.10$ && 1.57$\pm 0.05$ && +0.126 & 65 & 0.004 && +0.131 & 14 & 0.024 && FOP/Keck\\
Kepler-100&& 5855$\pm 40$ && 3.98$\pm 0.15$ && 1.49$\pm 0.06$ && +0.061 & 41 & 0.005 && +0.063 & 11 & 0.015 && Keck\\
Kepler-130&& 5958$\pm 37$ && 4.41$\pm 0.15$ && 1.78$\pm 0.11$ && -0.199 & 49 & 0.005 && -0.195 & 11 & 0.022 && Keck\\
\enddata
	
\end{deluxetable}

\begin{deluxetable}{lcccccccccccccc}
\tablecolumns{15}
\tablewidth{0pt}
\rotate
\tablecaption{Abundances\label{tab:abs}}
\tablehead{
	\colhead{Element}&
	\colhead{}&
	\colhead{Kepler-20}&
	\colhead{}&
	\colhead{Kepler-21}&
	\colhead{}&
	\colhead{Kepler-22}&
	\colhead{}&
	\colhead{Kepler-37}&
	\colhead{}&
	\colhead{Kepler-68}&
	\colhead{}&
	\colhead{Kepler-100}&
	\colhead{}&
	\colhead{Kepler-130}
	}
\startdata
$\log N{\mathrm{(Li)}}$ && $<1.10$ && $<1.20$ && $<1.00$ && $<0.85$ && 2.05 $\pm0.09$ && 2.21 $\pm0.08$ && 2.02 $\pm0.08$ \\
$\mathrm{[C/H]}$  && +0.02$\pm0.06$ && -0.02$\pm0.05$ && -0.24$\pm0.05$ &&-0.37$\pm0.03$ &&+0.08$\pm0.05$ && +0.11$\pm 0.08$ && -0.25$\pm 0.07$ \\
$\mathrm{[O/H]}$  && -0.04$\pm0.08$ && -0.05$\pm0.07$ && -0.19$\pm0.05$ &&-0.21$\pm0.07$ &&+0.09$\pm0.04$ && +0.19$\pm 0.10$ && +0.10$\pm 0.09$ \\
$\mathrm{[Na/H]}$ && +0.03$\pm0.03$ && -0.08$\pm0.05$ && -0.30$\pm0.03$ &&-0.33$\pm0.04$ &&+0.13$\pm0.03$ && +0.19$\pm 0.04$ && -0.24$\pm 0.03$ \\
$\mathrm{[Mg/H]}$ && +0.10$\pm0.04$ && -0.06$\pm0.04$ && -0.23$\pm0.03$ &&-0.28$\pm0.05$ &&+0.15$\pm0.03$ && +0.13$\pm 0.05$ && -0.14$\pm 0.04$ \\
$\mathrm{[Al/H]}$ && +0.08$\pm0.02$ &&    \nodata     && -0.21$\pm0.02$ &&-0.25$\pm0.04$ &&+0.20$\pm0.02$ && +0.16$\pm 0.03$ && -0.18$\pm 0.04$ \\
$\mathrm{[Si/H]}$ && +0.04$\pm0.02$ && -0.05$\pm0.02$ && -0.24$\pm0.02$ &&-0.32$\pm0.02$ &&+0.13$\pm0.01$ && +0.13$\pm 0.04$ && -0.15$\pm 0.02$ \\
$\mathrm{[S/H]}$  &&    \nodata     && -0.13$\pm0.05$ &&    \nodata     &&-0.37$\pm0.03$ &&+0.11$\pm0.04$ && +0.11$\pm 0.06$ && -0.33$\pm 0.05$ \\
\enddata

\end{deluxetable}

\begin{deluxetable}{lcrcccrcccrcccc}
\tablecolumns{15}
\tablewidth{0pt}
\tablecaption{Carbon Abundances\label{tab:carb}}
\tablehead{
	\colhead{Star}&
	\colhead{}&
	\colhead{[\ion{C}{1}/H]}&
	\colhead{}&
	\colhead{$\sigma$}&
	\colhead{}&
	\colhead{[C$_{2}$/H]}&
	\colhead{}&
	\colhead{$\sigma$}&
	\colhead{}&
	\colhead{$<$[C/H]$>$\tablenotemark{a}}&
	\colhead{}&
	\colhead{$N$}&
	\colhead{}&
	\colhead{$\sigma_{\mu}$}
	}
\startdata
Kepler-20 && +0.01 && 0.03 && +0.04 && 0.01  && +0.02 && 5 && 0.01 \\
Kepler-21 && -0.02 && 0.02 &&\nodata&&\nodata&& -0.02 && 4 && 0.01\\
Kepler-22 && -0.23 && 0.02 && -0.26 && 0.01  && -0.24 && 6 && 0.01 \\
Kepler-37 && -0.21 && 0.09 && -0.37 && 0.01  && -0.37 && 2 && 0.01 \\
Kepler-68 && +0.08 && 0.03 && +0.10 && 0.02  && +0.08 && 6 && 0.01 \\
Kepler-100&& +0.14 && 0.07 && +0.05 && 0.02  && +0.11 && 6 && 0.03 \\
Kepler-130&& -0.24 && 0.01 && -0.28 && 0.04  && -0.25 && 6 && 0.01 \\
\enddata

\tablenotetext{a}{Final adopted abundance.}

\end{deluxetable}

\begin{deluxetable}{lcrcrcccrcccr}
\tablecolumns{13}
\tablewidth{0pt}
\tablecaption{Oxygen Abundances\label{tab:oxy}}
\tablehead{
	\colhead{Star}&
	\colhead{}&
	\colhead{[\ion{O}{1}]}&
	\colhead{}&
	\multicolumn{7}{c}{\ion{O}{1} Triplet}&
	\colhead{}&
	\colhead{[O/H]\tablenotemark{a}}\\
	\cline{5-11}\\
	\colhead{}&
	\colhead{}&
	\colhead{[O/H]}&
	\colhead{}&
	\colhead{[O/H]$_{\mathrm{LTE}}$}&
	\colhead{}&
	\colhead{$\sigma$}&
	\colhead{}&
	\colhead{[O/H]$_{\mathrm{NLTE}}$}&
	\colhead{}&
	\colhead{$\sigma$}&
	\colhead{}&
	\colhead{}
	}
\startdata
Kepler-20 && -0.04 && -0.03 && 0.02 && -0.01 && 0.02  && -0.04 \\
Kepler-21 && -0.05 && +0.15 && 0.03 &&  0.00 && 0.00  && -0.05 \\
Kepler-22 &&\nodata&& -0.21 && 0.02 && -0.19 && 0.02  && -0.19 \\
Kepler-37 && -0.21 && -0.28 && 0.02 &&\nodata&&\nodata&& -0.21 \\
Kepler-68 &&\nodata&& +0.12 && 0.01 && +0.09 && 0.00  && +0.09 \\
Kepler-100&&\nodata&& +0.29 && 0.11 && +0.19 && 0.08  && +0.19 \\
Kepler-130&&\nodata&& +0.14 && 0.08 && +0.10 && 0.07  && +0.10
\enddata

\tablenotetext{a}{Final adopted abundance.}

\end{deluxetable}

\begin{deluxetable}{lcccccccccccc}
\tablecolumns{13}
\tablewidth{0pt}
\rotate
\tabletypesize{\footnotesize}
\tablecaption{Abundance Sensitivities\label{tab:sensi}}
\tablehead{
	\colhead{Species}&
	\colhead{}&
	\multicolumn{3}{c}{Kepler-20}&
	\colhead{}&
	\multicolumn{3}{c}{Kepler-21}&
	\colhead{}&
	\multicolumn{3}{c}{Kepler-22}\\
	\cline{3-5} \cline{7-9} \cline{11-13}\\
	\colhead{}&
	\colhead{}&
	\colhead{$\Delta T_{\mathrm{eff}}$}&
	\colhead{$\Delta \log g$}&
	\colhead{$\Delta \xi$}&
	\colhead{}&
	\colhead{$\Delta T_{\mathrm{eff}}$}&
	\colhead{$\Delta \log g$}&
	\colhead{$\Delta \xi$}&
	\colhead{}&
	\colhead{$\Delta T_{\mathrm{eff}}$}&
	\colhead{$\Delta \log g$}&
	\colhead{$\Delta \xi$}\\
	\colhead{}&
	\colhead{}&
	\colhead{($\pm150$ K)}&
	\colhead{($\pm0.25$ dex)}&
	\colhead{($\pm0.30 {\mathrm km s^{-1}}$)}&
	\colhead{}&
	\colhead{($\pm150$ K)}&
	\colhead{($\pm0.25$ dex)}&
	\colhead{($\pm0.30 {\mathrm km s^{-1}}$)}&
	\colhead{}&
	\colhead{($\pm150$ K)}&
	\colhead{($\pm0.25$ dex)}&
	\colhead{($\pm0.30 {\mathrm km s^{-1}}$)}
	}
\startdata
\ion{Fe}{1} && $\pm0.09$ & $\pm0.01$ & $\mp0.05$ && $\pm0.10$ & $\pm0.01$ & $\mp0.02$ && $\pm0.01$ & $\pm0.09$ & $\mp0.02$ \\    
\ion{Fe}{2} && $^{-0.07}_{+0.10}$ & $^{+0.13}_{-0.10}$ & $\mp0.05$ && $\mp0.02$ & $\pm0.10$ & $\mp0.05$ && $\pm0.11$ & $\mp0.06$ & $\mp0.03$ \\      
\ion{C}{1}  && $^{-0.09}_{+0.13}$ & $^{+0.10}_{-0.07}$ & $\mp0.01$ && $\mp0.08$ & $\pm0.09$ & $\mp0.01$ && $\mp0.10$ & $\pm0.09$ & $\mp0.01$ \\    
\ion{O}{1}  && $^{-0.15}_{+0.19}$ & $^{+0.09}_{-0.05}$ & $\mp0.02$ && $\mp0.11$ & $\pm0.07$ & $\mp0.05$ && $\mp0.15$ & $\pm0.08$ & $\mp0.02$ \\      
\ion{Na}{1} && $\pm0.10$ & $\mp0.05$ & $\mp0.03$ && $\pm0.07$ & $\mp0.02$ & $\mp0.02$ && $\pm0.09$ & $\mp0.03$ & $\mp0.02$ \\    
\ion{Mg}{1} && $\pm0.07$ & $\mp0.03$ & $^{-0.09}_{+0.05}$ && $\pm0.07$ & $\mp0.02$ & $\mp0.03$ && $\pm0.08$ & $\mp0.03$ & $\mp0.02$ \\     
\ion{Al}{1} && $\pm0.08$ & $\mp0.02$ & $\mp0.02$ &&  \nodata  &  \nodata  &  \nodata  && $\pm0.08$ & $\mp0.01$ & $\mp0.01$ \\      
\ion{Si}{1} && $\pm0.02$ & $\pm0.02$ & $\mp0.02$ && $\pm0.05$ & $\mp0.01$ & $\mp0.02$ && $\pm0.01$ & $\pm0.02$ & $\mp0.01$ \\
\enddata

\end{deluxetable}

\begin{deluxetable}{lllllllllllll}
\tablecolumns{13}
\rotate
\footnotesize
\tablewidth{0pt}
\tablecaption{Comparison of Stellar Parameters \label{tab:lit_params}}
\tablehead{
    \colhead{Star}&
    \colhead{}&
    \colhead{\teff}&
    \colhead{}&
    \colhead{$\log g$}&
    \colhead{}&
    \colhead{[Fe/H]}&
    \colhead{}&
    \colhead{Method}&
    \colhead{}&
    \colhead{Reference}&
    \colhead{}&
    \colhead{Comment}
    }
\startdata
Kepler-20 && 5514$\pm 25$  && 4.44$\pm 0.12$   && +0.06$\pm 0.06$ && MOOG      && this study                  &&  \\
          && 5455$\pm 100$ && 4.4$\pm 0.1$     && +0.01$\pm 0.04$ && SME       && \citet{2012ApJ...749...15G} &&  \\
          && 5466$\pm 93$  && 4.443$\pm 0.075$ && +0.02$\pm 0.04$ && SME       && \citet{2012Natur.482..195F} &&  \\
          && 5547$\pm 50$  && 4.55$\pm 0.10$   && +0.04$\pm 0.08$ && SPC       && \citet{2014Natur.509..593B} &&  \\

Kepler-21 && 6177$\pm 42$  && 3.99$\pm 0.15$   && -0.08$\pm 0.07$ && MOOG      && this study                  &&  \\
          && 6131$\pm 44$  && 4.0$\pm 0.1$     && -0.15$\pm 0.06$ && Multiple  && \citet{2012ApJ...746..123H} && 1\\
          && 6148$\pm 111$ && 3.94$\pm 0.21$   && -0.19$\pm 0.21$ && ROTFIT*   && \citet{2013MNRAS.434.1422M} && 2\\
          && 6409$\pm 74$  && 4.43$\pm 0.12$   && -0.03$\pm 0.06$ && ARES/MOOG && \citet{2013MNRAS.434.1422M} &&  \\
\enddata
\end{deluxetable}

\begin{deluxetable}{lcrcrcrcrcrcrcrcr}
\tablecolumns{17}
\tablewidth{0pt}
\tablenum{8}
\rotate
\tablecaption{Abundance Slopes with \tc\label{tab:slopes}}
\tablehead{
     \colhead{}&
     \colhead{}&
     \multicolumn{7}{c}{Unweighted Slope\tablenotemark{a}}&
     \colhead{}&
     \multicolumn{7}{c}{Weighted Slope}\\
     \cline{3-9} \cline{11-17}\\
     \colhead{Star}&
     \colhead{}&
     \colhead{w/Mn}&
     \colhead{}&
     \colhead{$\sigma$}&
     \colhead{}&
     \colhead{w/o Mn}&
     \colhead{}&
     \colhead{$\sigma$}&
     \colhead{}&
     \colhead{w/Mn}&
     \colhead{}&
     \colhead{$\sigma$}&
     \colhead{}&
     \colhead{w/o Mn}&
     \colhead{}&
     \colhead{$\sigma$}
     }
\startdata
Kepler-20  && +10.27 && 3.67 && +10.03 && 3.98 &&  +9.40 && 3.59 &&  +9.12 && 3.88  \\
Kepler-21  &&  -8.11 && 11.4 && -14.85 && 8.78 &&  -5.03 && 11.8 && -13.23 && 8.57  \\
Kepler-22  && +16.41 && 4.75 && +12.70 && 2.69 && +15.58 && 4.44 && +12.89 && 2.98  \\
Kepler-37  && +12.74 && 4.84 && +10.40 && 4.47 && +12.83 && 5.00 && +11.02 && 4.81  \\
Kepler-68  && +10.06 && 3.18 &&  +9.30 && 3.34 && +10.35 && 3.29 &&  +9.57 && 3.49  \\
Kepler-100 && -15.25 && 6.51 && -18.22 && 6.15 && -13.24 && 6.66 && -16.10 && 6.34  \\
Kepler-130 && +12.19 && 5.48 &&  +8.39 && 3.90 && +12.65 && 5.77 &&  +8.64 && 4.17  \\
\enddata

\tablenotetext{a}{all slopes are $\times 10^{-5}$ dex K$^{-1}$}

\end{deluxetable}

\begin{deluxetable}{lccccccccccccccccccccccccccc}
\tablecolumns{28}
\tablewidth{0pt}
\tablenum{9}
\rotate
\tabletypesize{\scriptsize}
\tablecaption{Lines Measured, Equivalent Widths, and Abundances\label{tab:lines}}
\tablehead{
     \colhead{}&
     \colhead{}&
     \colhead{$\lambda$}&
     \colhead{}&
     \colhead{$\chi$}&
     \colhead{}&
     \colhead{}&
     \colhead{}&
     \multicolumn{2}{c}{Kepler-20}&
     \colhead{}&
     \multicolumn{2}{c}{Kepler-21}&
     \colhead{}&
     \multicolumn{2}{c}{Kepler-22}&
     \colhead{}&
     \multicolumn{2}{c}{Kepler-37}&
     \colhead{}&
     \multicolumn{2}{c}{Kepler-68}&
     \colhead{}&
     \multicolumn{2}{c}{Kepler-100}&
     \colhead{}&
     \multicolumn{2}{c}{Kepler-130}\\
     \cline{9-10} \cline{12-13} \cline{15-16} \cline{18-19} \cline{21-22} \cline{24-25} \cline{27-28}\\
     \colhead{Ion}&
     \colhead{}&
     \colhead{(\AA)}&
     \colhead{}&
     \colhead{(eV)}&
     \colhead{}&
     \colhead{$\log \mathrm{gf}$}&
     \colhead{}&
     \colhead{EW}&
     \colhead{$\log N$}&
     \colhead{}&
     \colhead{EW}&
     \colhead{$\log N$}&
     \colhead{}&
     \colhead{EW}&
     \colhead{$\log N$}&
     \colhead{}&
     \colhead{EW}&
     \colhead{$\log N$}&
     \colhead{}&
     \colhead{EW}&
     \colhead{$\log N$}&
     \colhead{}&
     \colhead{EW}&
     \colhead{$\log N$}&
     \colhead{}&
     \colhead{EW}&
     \colhead{$\log N$}
     }
\startdata
\ion{C}{1}   && 5052.17 && 7.68 && -1.304 && 26.1 & 8.45 && 52.5 & 8.41 && 16.6 & 8.17 && 12.9 & 8.13 && 39.8 & 8.51   && 45.9 & 8.51         && 27.9 & 8.23 \\
             && 5380.34 && 7.68 && -1.615 && 14.6 & 8.44 && 34.0 & 8.42 && 10.7 & 8.27 &&  8.4 & 8.25 && 27.1 & 8.58   && 33.2 & 8.59         && 15.1 & 8.20 \\
             && 6587.61 && 8.54 && -1.021 &&  8.7 & 8.36 && 26.0 & 8.38 &&  7.2 & 8.26 &&\nodata&\nodata&& 20.5 & 8.56 && 26.4 & 8.61         && 10.5 & 8.16 \\ 
             && 7111.47 && 8.64 && -1.074 &&\nodata&\nodata&& 19.0 & 8.34 &&\nodata&\nodata && \nodata &\nodata && 14.4 & 8.51 && 13.8 & 8.36 &&  8.7 & 8.21 \\
             && 7113.18 && 8.65 && -0.762 &&\nodata&\nodata&&\nodata&\nodata&& 14.1 & 8.45 && 11.2 & 8.47 && 30.5 & 8.66 && 36.9 & 8.69       && 14.7 & 8.18 \\
\ion{O}{1}   && 6300.30\tablenotemark{a} && 0.00 && -9.720 &&  6.6 & 8.64 &&  5.1 & 8.63 &&\nodata&\nodata && 4.5 & 8.47 &&\nodata &\nodata && \nodata & \nodata &&\nodata &\nodata \\
             && 7771.94 && 9.15 &&  0.369 && 51.0 & 8.81 &&124.9 & 9.03 && 47.4 & 8.64 && 33.8 & 8.56 && 83.9 & 8.93 &&104.9 & 9.17 && 89.6 & 8.92 \\
             && 7774.17 && 9.15 &&  0.223 && 47.1 & 8.88 &&115.5 & 9.06 && 43.2 & 8.71 && 28.8 & 8.57 && 76.8 & 8.98 && 94.8 & 9.17 && 88.8 & 9.06 \\
\enddata

\tablenotetext{a}{Not corrected for Ni blend}
\tablenotetext{b}{Not corrected for hfs}

\end{deluxetable}

\begin{deluxetable}{lccccccccccccccc}
\tablecolumns{16}
\tablewidth{0pt}
\tablenum{10}
\rotate
\tabletypesize{\scriptsize}
\tablecaption{Solar Lines Measured, Equivalent Widths, and Abundances\label{tab:slines}}
\tablehead{
     \colhead{}&
     \colhead{}&
     \colhead{$\lambda$}&
     \colhead{}&
     \colhead{$\chi$}&
     \colhead{}&
     \colhead{}&
     \colhead{}&
     \multicolumn{2}{c}{Kepler-20, Kepler-21, Kepler-22}&
     \colhead{}&
     \multicolumn{2}{c}{Kepler-37, Kepler-68}&
     \colhead{}&
     \multicolumn{2}{c}{Kepler-100, Kepler-130} \\
     \cline{9-10} \cline{12-13} \cline{15-16} \\
     \colhead{Ion}&
     \colhead{}&
     \colhead{(\AA)}&
     \colhead{}&
     \colhead{(eV)}&
     \colhead{}&
     \colhead{$\log \mathrm{gf}$}&
     \colhead{}&
     \colhead{EW$_{\odot}$}&
     \colhead{$\log N_{\odot}$}&
     \colhead{}&
     \colhead{EW$_{\odot}$}&
     \colhead{$\log N_{\odot}$}&
     \colhead{}&
     \colhead{EW$_{\odot}$}&
     \colhead{$\log N_{\odot}$}
     }
\startdata
\ion{C}{1}   && 5052.17 && 7.68 && -1.304 && 31.7 & 8.41 && 32.3 & 8.43 && 33.2 & 8.46 \\
             && 5380.34 && 7.68 && -1.615 && 19.5 & 8.44 && 20.2 & 8.46 && 19.4 & 8.45 \\
             && 6587.61 && 8.54 && -1.021 && 14.1 & 8.42 && 16.4 & 8.51 && 13.1 & 8.39 \\ 
             && 7111.47 && 8.64 && -1.074 &&  9.4 & 8.36 && 13.3 & 8.54 && 10.5 & 8.42 \\
             && 7113.18 && 8.65 && -0.762 && 23.3 & 8.56 && 24.7 & 8.60 && 21.9 & 8.54 \\
\ion{O}{1}   && 6300.30\tablenotemark{a} && 0.00 && -9.720&& 5.6 & 8.68 &&\nodata&\nodata \\
             && 7771.94 && 9.15 &&  0.369 && 71.2 & 8.86 && 68.5 & 8.81 && 65.6 & 8.80 \\
             && 7774.17 && 9.15 &&  0.223 && 64.8 & 8.90 && 61.7 & 8.85 && 58.7 & 8.83 \\
\enddata

\tablenotetext{a}{Not corrected for Ni blend}
\tablenotetext{b}{Not corrected for hfs}

\end{deluxetable}

\end{document}